\documentclass[aps,twocolumn,prd,dvipsnames,usenames,floatfix,superscriptaddress,nofootinbib,longbibliography]{revtex4-2}
\pdfoutput=1

\usepackage[utf8]{inputenc}
\usepackage{units}
\usepackage[english]{babel}
\usepackage{xspace}
\usepackage{amsmath} 
\usepackage{amssymb} 
\usepackage{graphicx}
\usepackage{hyperref}
\usepackage{color}

\usepackage{xspace}
\newcommand*{\ie}{i.e., }
\newcommand*{\eg}{e.g., }
\newcommand*{\fig}{fig.\@\xspace}
\newcommand*{\eq}{eq.\@\xspace}
\newcommand*{\eqs}{eqs.\@\xspace}

\newcommand*{\rhs}{r.h.s.\@\xspace}
\newcommand*{\lhs}{l.h.s.\@\xspace}

\renewcommand{\L}{\mathcal{L}}

\renewcommand{\d}{\partial}

\def\e{{\rm e}}

\usepackage{xifthen}
\newcommand*\diff{\mathrm{d}} 
\newcommand*\ldiff[2][]{ \ifthenelse{\isempty{#1}}{ \diff #2}{\diff^#1#2} \,} 

\begin{abstract}
	
	We propose a model for combining the Standard Model (SM) with gravity. It relies on a non-minimal coupling of the Higgs field to the Ricci scalar and on the Palatini formulation of gravity. Without introducing any new degrees of freedom in addition to those of the SM and the graviton, this scenario achieves two goals.  First, it generates the electroweak symmetry breaking by a non-perturbative gravitational effect. In this way, it does not only address the hierarchy problem but opens up the possibility to calculate the Higgs mass. Second, the model incorporates inflation at energies below the onset of strong-coupling of the theory. Provided that corrections due to new physics above the scale of inflation are not unnaturally large, we can relate inflationary parameters to data from collider experiments.
	
\end{abstract}

\begin{document}
		\title{Standard Model Meets Gravity: Electroweak Symmetry Breaking and Inflation}
	\author{Mikhail Shaposhnikov} 
	\email{mikhail.shaposhnikov@epfl.ch}
	\author{Andrey Shkerin} 
	\email{andrey.shkerin@epfl.ch}
	\author{Sebastian Zell}
	\email{sebastian.zell@epfl.ch}
	\affiliation{Institute of Physics, Laboratory for Particle Physics and Cosmology,
		\'Ecole Polytechnique F\'ed\'erale de Lausanne, CH-1015 Lausanne, Switzerland}

	\maketitle

	\section{Introduction} The results of LHC have been very exciting. First, it has found the last missing particle of the Standard Model (SM), the Higgs boson \cite{Aad:2012tfa,Chatrchyan:2012xdj}. Secondly, it has significantly constrained physics beyond the SM. In many scenarios, the existence of new particles close to the electroweak scale is now excluded. This gives a significant motivation to study the proposal that no new degrees of freedom exist anywhere above the weak scale $M_F\sim 10^2\:$GeV (see e.g., \cite{Shaposhnikov:2007nj}).  Such a situation is self-consistent,  since with the measured values of its parameters, the SM is a valid quantum field theory until the Landau poles in the Higgs self-interaction and the hyper-charge gauge interaction, which appear at exponentially large energies, well above another fundamental scale of Nature  -- the Planck mass $M_P=2.44\cdot 10^{18}\:$GeV.\footnote
	{Depending on the masses of the top quark and of the Higgs boson, the Higgs self-coupling can become negative at energy scales between $10^{8}\:$GeV and $M_P$ and thereby give rise to another, deeper minimum of the Higgs potential \cite{Degrassi:2012ry, Buttazzo:2013uya}. Whether this happens or not is an open question, given the uncertainties in the determination of the top quark Yukawa coupling; see \cite{Bezrukov:2014ina} for a review. But even if our current vacuum is metastable, the validity of the SM is not spoiled, since its lifetime exceeds the age of the Universe by many orders of magnitude \cite{Andreassen:2017rzq}.} 
	
	Other experimental and observational data that calls for new physics, such as dark matter, neutrino oscillations and baryon asymmetry of the Universe,  does not require the presence of any new particle populating the desert between the Fermi and the Planck scales, either.\footnote
	{For example, the Neutrino Minimal Standard Model ($\nu$MSM) \cite{Asaka:2005an,Asaka:2005pn}, whose particle content is extended compared to that of the SM only by three Majorana neutrinos with masses below $M_F$, may account for all these  phenomena in a unified way (for a review see \cite{Boyarsky:2009ix}).}
	Moreover, the existence of new heavy particles (such as leptoquarks of Grand Unified Theories) leads to the celebrated  problem of the stability of the Higgs mass against radiative corrections coming from loops with these superheavy states \cite{Gildener:1976ai}. If there are no such particles all together, the hierarchy problem as a concern about the sensitivity of low-energy parameters to high-energy physics below the Planck scale disappears \cite{Vissani:1997ys, Shaposhnikov:2007nj, Farina:2013mla}.  Another aspect of the problem, however, remains, and it is centered around the question why the electroweak scale is so much smaller than the Planck scale. This is one of the issues that we shall address in the present work.
	
	If we have only the SM (or $\nu$MSM) degrees of freedom all the way up to the Planck scale, the question arises:  ``How does the SM merge with gravity?''  In this paper we show that the \textit{conformally-invariant}  (at the classical level) SM coupled to gravity in the Palatini formulation  with non-minimal  interaction between the Higgs field and the gravitational Ricci scalar has a number of remarkable properties indicating, perhaps, that this is a step in the right direction. The Lagrangian of the model reads:
	\begin{equation} 
	\label{theory}
	\L=-\dfrac{M_P^2}{2}R-\xi H^\dagger H R+ \L_{\rm SM}|_{m_\text{H,tree}\to 0}\;,
	\end{equation}
	where $R$ is the Ricci scalar, $H$ is the Higgs field, $\xi>0$ is the strength of its non-minimal coupling to gravity, $\L_{\rm SM}$ is the SM Lagrangian, and $m_\text{H,tree}$ is the tree-level Higgs mass.
	
	We start from the well known facts about different sectors of this theory. In the Palatini formulation of gravity \cite{Palatini1919, Einstein1925}, the metric $g_{\mu\nu}$ and the symmetric affine connection $\Gamma^\alpha_{\beta \gamma}$ are treated as independent variables. In spite of the larger number of field components as compared to metric gravity, the number of physical propagating degrees of freedom -- two of the massless graviton -- is the same in both theories. In the absence of the non-minimal coupling, $\xi=0$, Palatini gravity is moreover exactly equivalent to the standard metric Einstein gravity.
	
	The particle physics sector of the theory is the SM with zero tree-level Higgs mass. It is well known that the Lagrangian $\L_{\rm SM}|_{m_\text{H,tree}\to 0}$ has an extra symmetry -- it is invariant under the group of conformal transformations. What is most important for us is that in this theory the Higgs mass is predictable \cite{Coleman:1973jx,Weinberg:1976pe,Linde:1977mm} (to be more precise, the ratio between the scalar and vector boson mass is computable). The easiest way to see that is to use the minimal subtraction scheme for removing the divergencies. Here the counter-terms are polynomials in the coupling constants \cite{tHooft:1973mfk}, and, if $m_\text{H,tree}=0$, no counter-term is needed for the mass renormalization, meaning that the Higgs mass can be expressed through other parameters of the theory. This is true even in the presence of gravity, because perturbative quantum gravity corrections can only contain inverse powers of $M_P$ \cite{tHooft:1974toh}. To put it in different words, the renormalization group $\beta$-function for the Higgs mass is zero if $m_\text{H,tree}=0$ \cite{tHooft:1973mfk}.\footnote{That this remains true in the presence of a non-minimal coupling $\xi$ is discussed in appendix A.}
	
	First, we are going to argue that electroweak symmetry breaking in the theory (\ref{theory}) can be induced by the non-perturbative semiclassical effect related to a singular gravitational-scalar instanton -- a solution to classical equations of motion of Euclidean gravity.\footnote
	{Note that the Coleman-Weinberg effective potential \cite{Coleman:1973jx} in the SM cannot lead to electroweak symmetry breaking with the experimental values of the Higgs self-coupling and top quark Yukawa coupling (see, e.g., \cite{Witten:1980ez,Froggatt:1995rt}).}
	This effect has been already discussed in metric gravity \cite{Shaposhnikov:2018xkv}, but there it was difficult to implement and required ad-hoc ingredients. We will show that in the Palatini formulation, the gravitational instanton can be realized significantly more simply and naturally.
	For large values of the non-minimal coupling $\xi$, we find that $M_F \propto M_P \exp{(-B)}$, where $B$ is the instanton action. The observed hierarchy of the Fermi and Planck scales requires $B\sim 30$, which we can easily achieve.
	
	Second, we will show that the very same choice of parameters leads to successful inflation. The role of the inflaton is played by the Higgs field \cite{Bezrukov:2007ep}. Due to the non-minimal coupling of the Higgs field to gravity, predictions of Higgs inflation in the Palatini formulation of gravity are different from those in the metric case \cite{Bauer:2008zj}. The prominent feature of this scenario is the increase of the energy scale $\Lambda$, at which tree-level unitarity is violated.\footnote
	{The scale $\Lambda$ only applies to scattering in a vacuum background. In metric Higgs inflation, the scale of unitarity violation is background-dependent and during inflation it lies above inflationary energies
			 \cite{Bezrukov:2010jz} (see also \cite{Escriva:2016cwl, Fumagalli:2017cdo}). In contrast, we expect in the Palatini case that the scale of unitarity violation does not increase in a non-trivial background (see appendix B).}
	In the original Higgs inflation, $\Lambda$ is of the order of $M_P/\xi$ and lies below inflationary scales
	\cite{Barbon:2009ya,Burgess:2009ea}.  
	On the one hand, this makes it impossible to determine the inflationary potential from the low-energy SM parameters unless the ``jumps'' of the coupling constants at the onset of the strong coupling regime happen to be very small \cite{Bezrukov:2010jz, Bezrukov:2014ipa}. 
		On the other hand, the low value of $\Lambda$ is expected to lead to a breakdown of perturbation theory during preheating \cite{Ema:2016dny, DeCross:2016cbs}. 
		In contrast, Palatini Higgs inflation gives $\Lambda = M_P/\sqrt{\xi}$ \cite{Bauer:2010jg}, which lies above inflationary energies. As discussed in more detail in \cite{SecondPaper}, this allows us to establish a connection between low- and high-energy parameters of the theory, provided that corrections due to new physics are not unnaturally large. Moreover, no strong coupling is expected to occur during preheating.
		It is important to recall that in our approach no new particles exist above the weak scale. Consequently, the violation of tree-level unitarity at the scale $\Lambda$ is due to a strong-coupling regime of the low-energy degrees of freedom.\footnote
		{Further discussions of this idea of ``self-healing'' \cite{Bezrukov:2010jz, Aydemir:2012nz} can be found in \cite{SecondPaper}. In the context of gravity, such a scenario was, \eg suggested in \cite{Weinberg:1980gg, Reuter:1996cp, Dvali:2010bf, Dvali:2010jz, Dvali:2011th}.}

\section{The model} We are interested in the Higgs-gravity sector of the model (\ref{theory}). The rest of the SM particles manifest themselves through RG running of the Higgs quartic coupling $\lambda$, which shapes the effective Higgs potential.
We neglect the running of $\xi$ (see appendix C).
 When we apply the unitary gauge for the Higgs field, $H=(0,h)^T/\sqrt{2}$, the relevant part of the Lagrangian reads
\begin{equation} \label{jordanFrame}
\L=-\dfrac{M_P^2 + \xi h^2}{2}R+\frac{1}{2}(\partial_\mu h)^2-\frac{\lambda}{4}h^4 \;.
\end{equation}
In order to make the kinetic term of $h$ canonical, we perform a Weyl transformation of the metric, 
\begin{equation}\label{Omega}
\hat{g}_{\mu\nu} = \Omega^2 g_{\mu\nu} \;, ~~~ \Omega^2 = 1 + \dfrac{\xi h^2}{M_P^2} \;,
\end{equation}
followed by the field redefinition \cite{Bauer:2008zj}
\begin{equation} \label{fieldTransformation}
h = \frac{M_P}{\sqrt{\xi}} \sinh\left(\frac{\sqrt{\xi}\chi}{M_P}\right) \,.
\end{equation}
Then Lagrangian (\ref{jordanFrame}) becomes
\begin{equation}\label{einsteinFrame}
\L=-\frac{M_P^2}{2} \hat{R} + \frac{1}{2}(\partial_\mu \chi)^2 -U(\chi) \,,
\end{equation}
and the scalar potential is given by
\begin{equation}\label{potential}
U(\chi) = \frac{\lambda M_P^4}{4 \xi^2}\left(\tanh\left(\frac{ \sqrt{\xi}\chi}{M_P}\right)\right)^{4} \,.
\end{equation}
Note that if we had worked in the metric formulation of the theory (\ref{theory}), we would have arrived at the same form (\ref{einsteinFrame}) of the Lagrangian but with a different potential $U(\chi)$. Thus, the non-equivalence of the Palatini and metric formalisms manifests itself as the difference in the self-interaction of the canonically normalized scalar field.

\section{Fermi scale}  Let us discuss how the model (\ref{theory}) can elegantly accommodate the non-perturbative mechanism of generation of the Fermi scale proposed in \cite{Shaposhnikov:2018xkv} and developed further in \cite{Shaposhnikov:2018jag,Shkerin:2019mmu}. Our starting point is the expectation value of $h$ in the path integral formalism:
\begin{equation}
\label{PathIntegral}
\langle h \rangle\sim\int Dh Dg_{\mu\nu} D\Gamma^\alpha_{\beta \gamma}\,  \: h\, \e^{-S_E} \;,
\end{equation}
where $S_E$ is the Euclidean action of the theory.\footnote
{Note that because of the presence of gravity, the Euclidean path integral in \eq (\ref{PathIntegral}) must be taken with caution \cite{Gibbons:1977zz,Gratton:1999ya}.}
We disregarded the rest of the SM degrees of freedom since they can only change the prefactor but not the exponential dependence in our subsequent result \eqref{Hierarchy}. 

Our goal is to study if the path integral \eqref{PathIntegral} possesses saddle points besides the trivial one at $\langle h\rangle=0$. To this end, we notice that by making the change of variable according to \eqs (\ref{Omega}), (\ref{fieldTransformation}), the expectation value can be written as
\begin{equation}
\label{PathIntegral3}
\langle h\rangle\sim\dfrac{M_P}{\sqrt{\xi}}\int D\chi D\hat{g}_{\mu\nu} D\Gamma^\alpha_{\beta \gamma} \, \e^{\frac{\sqrt{\xi}\chi}{M_P}-S_E} \;.
\end{equation}
Here we used that $h>0$ (and, correspondingly, $\chi>0$) in the unitary gauge. Next, we integrate out the connection field. To this end, we split $\Gamma^\alpha_{\beta \gamma}$ into the Levi-Civita part and the contorsion tensor $C^\alpha_{\beta \gamma}$. This does not change the integration measure in \eqref{PathIntegral3}: $\int D\Gamma=\int DC$. The contorsion enters only the gravitational part of the action $S_E$ and reads (see, e.g., \cite{Dadhich:2010xa}):
\begin{equation}\label{Ricci}
\begin{split}
\int  & \diff^4x\sqrt{g}g^{\mu\nu}R_{\mu\nu}(\Gamma)=\int\diff^4x\sqrt{g}g^{\mu\nu} \left( R_{\mu\nu}(g) \right. \\
& \left.-\nabla_\mu C^\rho_{\rho\nu}+\nabla_\rho C^\rho_{\mu\nu}+C^\lambda_{\mu\nu}C^\rho_{\rho\lambda}-C^\lambda_{\rho\nu}C^\rho_{\mu\lambda} \right) \;.
\end{split}
\end{equation}
We see that the path integral over the contorsion is Gaussian. Hence it can be evaluated exactly by solving the equations of motion for $C^\alpha_{\beta \gamma}$ and plugging the result back in the action. Varying \eqref{Ricci} with respect to $C^\alpha_{\beta \gamma}$ gives the condition of vanishing contorsion: $C^\alpha_{\beta \gamma}=0$, where we used that in Palatini gravity the connection is assumed to be symmetric: $\Gamma^\alpha_{\beta \gamma} = \Gamma^\alpha_{\gamma \beta}$. Thus, the metric and Palatini formulations are equivalent in the absence of non-minimal coupling and \eq \eqref{PathIntegral3} becomes
\begin{equation}
\label{PathIntegral2}
\langle h\rangle\sim\dfrac{M_P}{\sqrt{\xi}}\int D\chi D\hat{g}_{\mu\nu} \, \e^{\frac{\sqrt{\xi}\chi}{M_P}-S_E} \;,
\end{equation}
where the Levi-Civita connection is now used in $S_E$. 
Note that both integration measures $D\chi$ and $ D\hat{g}_{\mu\nu}$ can contain non-trivial factors, which we show and discuss in appendix D. We will argue shortly that all of them are inessential for our purposes as long as we are interested in the leading-order exponential contribution to \eq \eqref{PathIntegral2}.

In \eq \eqref{PathIntegral2}, it is natural to expect the term $\sqrt{\xi}\chi/M_P$ to be included in the determination of the saddle point.\footnote{The same approach is used, e.g., in the discussion of confinement in gauge theories \cite{Polyakov:1978vu} and of multiparticle production \cite{Khlebnikov:1990ue}.}  As in \cite{Shaposhnikov:2018xkv, Shaposhnikov:2018jag,Shkerin:2019mmu}, our subsequent analysis is based on this assumption. If it holds, then the dominant contribution to the path integral is provided by extrema of 
	\begin{equation}\label{ActionMod}
	\mathcal{B}=\int \! \diff^4 x \left(-\frac{\sqrt{\xi}\chi(x)}{M_P} \delta^{(4)}(x)+ \sqrt{\hat{g}_E}\:\mathcal{L}_E\right) \;.
	\end{equation}
The subscript $E$ refers to the Euclidean signature. We see that the Lagrangian is supplemented by an instantaneous source and we used translational invariance of the theory to evaluate the latter at the origin. The corresponding saddle-point approximation gives
\begin{equation}\label{Hierarchy}
\langle h\rangle\sim\dfrac{M_P}{\sqrt{\xi}} e^{-B} \;,
\end{equation}
where $B$ is the value of $\mathcal{B}$, evaluated at a suitable Euclidean classical configuration of the fields $\chi$ and $\hat{g}_{\mu\nu}$.  The approximation \eqref{Hierarchy} only holds if $B$ is large, since solely in this case fluctuations above the classical background are suppressed. Clearly, the same requirement naturally leads to a hierarchy between the scales $M_P$ and $M_F$. One must show that $\mathcal{B}$ possesses extrema such that the resulting action is large but finite.

The extremum points of $\mathcal{B}$ are found by varying it with respect to $\chi$ and $\hat{g}_{\mu\nu}$. This yields Euclidean equations of motion supplemented by the instantaneous  source of $\chi$. Since the point source preserves the $O(4)$-symmetry of the theory, we specialize to spherically-symmetric solutions. The assumption that solutions of maximal symmetry minimize an Euclidean action is commonly used in studies of Euclidean gravity \cite{Coleman:1980aw} (see also \cite{Hawking:1998bn, Garriga:1998tm, Vilenkin:1998pp, Giddings:1987cg}), although the proof of it is only known for scalar field theories in flat space  \cite{Coleman:1977th,Blum:2016ipp}.
Thus, we choose the following ansatz for the metric: 
\begin{equation} \label{metric}
\diff s^2=f(r)^2\diff r^2+r^2\diff \Omega_3^2 \;,
\end{equation}
where $\diff \Omega_3$ is the line element on a unit 3-sphere and $f$ is a function of the radial coordinate $r$. The  total action \eqref{ActionMod} becomes
		\begin{equation} \label{actionSymmetric}
	\mathcal{B} = \int_0^\infty \diff r \left(\frac{\sqrt{\xi}\chi'(r)}{M_P} + 2 \pi^2 r^3 f\:\mathcal{L}_E\right) \;.
\end{equation}

Before determining extrema of $\mathcal{B}$, let us discuss why we expect subleading contribution, \eg from non-trivial measure factors in the path integral, to be small as compared to the saddle-point result. For this purpose, we implement a scaling argument to extract explicitly the large parameter in the total action \eqref{actionSymmetric}. We shall restrict ourselves to the spherically-symmetric ansatz \eqref{metric}. Moreover, we assume that $f\to 0$ as $r\to 0$;\footnote{This allows us to neglect the second term in the scalar curvature $R\sim(-f+f^3+rf')/(r^2f^3)$.} the actual solution fulfills this property: see subsequent discussion below \eqs \eqref{EoMSD1}, \eqref{EoMSD2}. With these simplifications, the change of variables
\begin{equation}\label{scaling}
r\to \xi^{1/6}r/M_P \;, ~~~ \chi\to M_P\chi \;, ~~~ f\to \xi^{-1/6}f
\end{equation} 
brings the total action to the form 
\begin{equation}\label{DimLessAction}
\mathcal{B}=\sqrt{\xi}\bar{\mathcal{B}} \;,
\end{equation}
where $\bar{\mathcal{B}}$ is a dimensionless action in which all terms are of the order of one. Thus, the instanton action $B$ is large as long as the non-minimal coupling $\xi$ is large. Moreover, all terms in the quadratic action for fluctuations are manifestly of the order of one, which gives us an argument that the corrections to the saddle-point result \eqref{Hierarchy} are subleading.

It turns out, however, that evaluating $\mathcal{B}$ in the theory \eqref{einsteinFrame} leads to an infinite action, caused by a divergent value of $\chi(0)$. Indeed, the equations of motion take the form
\begin{align}
& \frac{r^3\chi'}{f}=-\frac{\sqrt{\xi}}{2\pi^2M_P} \;,\label{EoMSD1}\\
& 6-6f^2-\frac{r^2\chi'^2}{M_P^2}=0 \;,\label{EoMSD2}
\end{align}
where for the moment we switched off the potential. One boundary condition for the instanton comes from asymptotic flatness: $\chi\to 0$, $f\to 1$ at $r\to\infty$; another one is due to the source: $f\to 0$ as $r\to 0$.
Together, they select a unique solution of \eqs \eqref{EoMSD1},\eqref{EoMSD2}. Its exact form is
\begin{align}
& f(r)=\sqrt{\frac{1}{1+\frac{\xi}{24\pi^4M_P^4r^4}}} \;, \\ 
& \chi(r)=\frac{\sqrt{\xi}}{2\pi^2M_P}\int_r^\infty\frac{f(r')dr'}{r'^3} \;.
\end{align}
We see that at $ r\to 0$
\begin{equation}\label{As}
f\approx \frac{2\sqrt{6}\pi^2M_P^2r^2}{\sqrt{\xi}} \;, ~~~ \chi\approx -\sqrt{6}M_P\log r \;.
\end{equation}
Thus, the scalar field diverges at the origin: $\chi(0)=\infty$ and the solution is not viable.
It is easy to check that allowing for non-zero potential $U(\chi)$ does not change the above results. Indeed, since on the solution both $r^3fU'(\chi)$ and $r^2 f^2 U(\chi)$ tend to $0$ at $r\to 0$, the short-distance asymptotics of the instanton are still given by \eqs (\ref{As}).

At this point, we must remember that our theory enters strong coupling at a finite energy scale. Therefore, its high-energy behavior is sensitive to the existence of higher-dimensional operators. We can use those to remove the unphysical UV-divergence of $\chi(0)$. As we shall show, it suffices to supplement Lagrangian \eqref{jordanFrame} by the operator (in the Lorentz signature)\footnote
{We choose this operator since the simplest option $ \propto -\delta(\d_\mu h)^4$ (with positive $\delta$) would violate positivity bounds \cite{Adams:2006sv, Herrero-Valea:2019hde}.}
\begin{equation}\label{OpDelta}
\delta\L_\delta=\dfrac{\delta}{M_P^8\Omega^8}\left(1+\frac{\delta}{\Omega^2}\right)(\partial_\mu h)^6 \,,
\end{equation}
 where $\delta>0$ and $\Omega$ is defined in \eq (\ref{Omega}). Clearly, $\delta\L_\delta$ does not introduce any new degrees of freedom. Hence, it fulfills our assumption about the absence of new particles above the weak scale. Moreover, the operator \eqref{OpDelta} is suppressed below the scale $\sim M_P/\sqrt{\xi}$ as long as $\delta \lesssim \xi^2$. Nevertheless, it is important to emphasize that our goal is not to discuss a possible UV-completion of the theory connected to the specific choice of the operator \eqref{OpDelta}. Instead, we want to demonstrate on a simple example how regularization at high energies can be achieved. The operator (\ref{OpDelta}) is not a unique option; other derivative operators produce the same effect on the instanton \cite{Shaposhnikov:2018xkv}. 	

After including the operator \eqref{OpDelta}, the Euclidean equations of motion become
\begin{equation}
\label{EoMs}
\begin{split}
& \scalebox{0.9}[1]{ $\d_r\left(\dfrac{r^3\chi'}{f}+\dfrac{6\delta r^3\chi'^5 G(\chi)}{M_P^8f^5}\right) - \dfrac{\delta r^3\chi'^6 G'(\chi)}{M_P^8f^5}$}\\
&~~~~~~~~~~~~~~~~~~~~~~~~~~~ \scalebox{0.9}[1]{ $ - r^3fU'(\chi) =-\dfrac{\sqrt{\xi}}{2 \pi^2 M_P}\delta(r) \;, $}\\
& \scalebox{0.9}[1]{$  6-6f^2+\dfrac{2r^2f^2U(\chi)}{M_P^2}- \dfrac{r^2\chi'^2}{M_P^2}-\dfrac{10\delta r^2\chi'^6 G(\chi)}{M_P^{10}f^4} =0$ }\;,
\end{split}
\end{equation}
where we defined $G(\chi)=1+\delta/\cosh^2\left(\sqrt{\xi}\chi/M_P\right)$. We shall solve these equations numerically with the boundary conditions discussed above. Close to the origin, however, they reduce to
\begin{align}
& \frac{6\delta r^3\chi'^5}{M_P^8f^5}=-\frac{\sqrt{\xi}}{2\pi^2M_P} \;, \label{EOMSD3} \\
&  6-\frac{10\delta r^2\chi'^6}{M_P^{10}f^4}=0 \;,\label{EOMSD4}
\end{align}
and this again can be solved analytically, yielding the short-distance asymptotics of the instanton:
\begin{align}
& f(r)\sim\delta^{\frac{1}{10}}\xi^{-\frac{3}{10}}(M_Pr)^{\frac{4}{5}} \;, \\
& \chi'(r)\sim\delta^{-\frac{1}{10}}\xi^{-\frac{1}{5}}M_P^{\frac{11}{5}}r^{\frac{1}{5}} \;. 
\end{align}
We see that $\chi(0)$ is finite now. It remains to check that one can achieve $B=\mathcal{O}(10)$.
To this end, we use \eq \eqref{EoMs} to determine $\delta$ as a function of $\xi$ in such a way that $\langle h\rangle=M_F$. The method for computing the instanton and the corresponding action is described in appendix E, and the result is shown in \fig \ref{fig:XiVsDelta}. For the values of $\xi$ admissible for inflation (see below), $\delta$ is of the same order of magnitude as $\xi^2$.

\begin{figure}[t]
	\begin{center}
		\center{\includegraphics[scale=0.7]{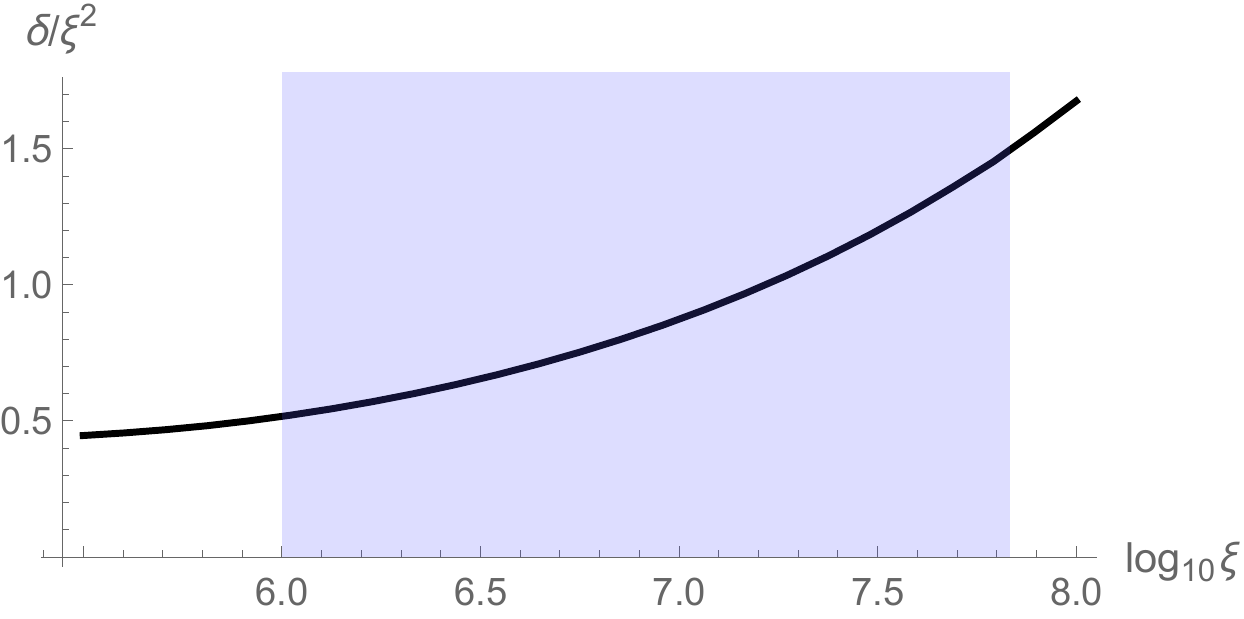}}
		\caption{Values of the non-minimal coupling $\xi$ and the coupling $\delta$ of the higher-order operator \eqref{OpDelta}, for which $B=\ln (M_P/(\sqrt{\xi}M_F))$. Admissible values of $\xi$ are within the blue area, the left bound coming from inflation and the right bound coming from top quark measurements.}
		\label{fig:XiVsDelta}
	\end{center}
\end{figure}

As discussed above, a necessary condition for the validity of the saddle-point result \eqref{Hierarchy} is $B\gg1$. In the theory without the higher-dimensional operator, the scaling argument \eqref{scaling} shows that the strength of the source term, controlled by the non-minimal coupling $\xi$, determines the instanton profile and appears as a common factor in $B$. Therefore, a value $\xi \gg 1$ automatically gives a large $B$, unlike in the scenario considered in \cite{Shaposhnikov:2018xkv}. In the theory with the operator \eqref{OpDelta}, the scaling argument remains applicable although it becomes more complicated. For completeness, we discuss it in appendix E. We postpone a more detailed study of subleading corrections to the instanton solution to future work.

Finally, one may wonder whether the instanton mechanism can also be realized in a simpler setup. For example, consider the theory with the same Lagrangian for $\chi$ but without gravity, in the flat background, \ie set $f=1$.  In this case, it follows from \eq \eqref{EOMSD3} that $\chi' \sim r^{-3/5}$ at small $r$, hence $\chi(0)$ is still finite. Furthermore, one can show that the instanton action is also finite. However, the strength of the source now is of order 1, \ie $\sqrt{\xi}$ disappears from the \rhs of the first line of \eq\eqref{EoMs}. Consequently, the instanton action is also of order 1 and the semiclassical approximation is no longer valid. Dynamical gravity and a large non-minimal coupling $\xi$ are, therefore, essential to make the action large. The same conclusion was reached in metric gravity \cite{Shaposhnikov:2018xkv,Shaposhnikov:2018jag,Shkerin:2019mmu}.

\section{Inflation}  The potential (\ref{potential}) gives rise to inflation at field values $\chi\gtrsim M_P/\sqrt{\xi}$ \cite{Bauer:2008zj}. The spectral tilt and tensor-to-scalar ratio are readily computed:
\begin{equation}\label{ns,r}
n_s=1-\dfrac{2}{N} \,, ~~~ r=\dfrac{2}{\xi N^2} \;,
\end{equation}
where $N$ is the number of e-foldings. In what follows, we take $N=50.9$ corresponding to $\xi \sim 10^7$ \cite{Rubio:2019ypq}.  The prediction for $n_s$ is essentially identical to the original scenario of Higgs inflation \cite{Bezrukov:2007ep}, but $r$ is suppressed by an additional power of $\xi$ \cite{Bauer:2008zj}.  One can use the normalization of the inflationary potential, extracted e.g., from the Planck data \cite{Akrami:2018odb}, to relate $\xi$ and $\lambda$:
\begin{equation}\label{Norm}
\xi=1.1\cdot 10^{10}\lambda \;.
\end{equation}
At this point, the question arises if the high-energy value of $\lambda$, which appears in \eq \eqref{Norm}, can be derived from the parameters of the SM measured at collider experiments. The relevant energy for the evaluation of the corresponding RG evolution is of the order of the top quark mass, $\mu = y_t M_P/\sqrt{\xi}$, where $y_t$ is the top Yukawa coupling and $y_t \approx 0.43$ at inflationary energies \cite{SecondPaper}. It lies below the scale $\Lambda = M_p/\sqrt{\xi}$, at which perturbation theory (defined on top of the low-energy vacuum) breaks down \cite{Bauer:2010jg}.\footnote
	{Also in the metric formalism, it is possible to increase $\Lambda$ by adding higher dimensional operators \cite{Germani:2010gm, Lerner:2010mq} or new degrees of freedom \cite{Giudice:2010ka}.}
	However, the separation of $\mu$ and $\Lambda$ is small and, moreover, $\lambda$, as evaluated within the SM, is close to zero and, therefore, susceptible to corrections. For this reason, the connection of low- and high-energy physics may break down if contributions of strongly-coupled physics at $\Lambda$ are unnaturally large \cite{SecondPaper}. But if this is not the case, inflationary parameters can be deduced from the low-energy data using the SM running of the relevant couplings.

To a good accuracy the running of $\lambda$ within the SM can be presented as
\begin{equation}\label{RunningLambda}
\lambda(\mu)=\lambda_0+b\ln^2\left(\dfrac{\mu}{qM_P}\right) \;.
\end{equation}
Here $q\lesssim 1$, $b\sim 10^{-5}$, and $\lambda_0\ll 1$ are functions of the parameters of the SM. Today, the largest uncertainty in their determination comes from measurements of $y_t$ \cite{Bezrukov:2014ina}. Plugging in $\lambda(\mu)$ in \eq (\ref{Norm}), we can determine $\xi$ as a function of $y_t$ measured at the weak scale.
For example, taking the conservative bound $m_t \gtrsim 170\,\text{GeV}$ \cite{Hoang:2014oea,Khachatryan:2015hba,Ravasio:2018lzi,Aaboud:2018zbu} on the top pole mass as an input, we get \cite{SecondPaper}
\begin{equation}\label{BoundOnXi1}
\xi < 6.8 \cdot 10^{7} \,.
\end{equation}
Thus, barring the above remark about corrections due to strong coupling at $\Lambda$, the lower bound on the top mass inferred from collider experiments leads to an upper bound on $\xi$. Improving precision in top quark measurements narrows down the window of admissible values of $\xi$.

Inflation itself provides a lower bound on $\xi$, as was already noticed in \cite{Rasanen:2017ivk}. It is given by \cite{SecondPaper}:
\begin{equation}\label{BoundOnXi2}
\xi> 1.0 \cdot 10^6 \,.
\end{equation}
Essentially, this constraint comes from the requirement that after plugging in  $\lambda(\mu)$ from \eq (\ref{RunningLambda}), the potential \eqref{potential} does not develop a stationary point below $\mu$.
If we take the intermediate value $\xi=10^7$ in between the bounds (\ref{BoundOnXi1}) and (\ref{BoundOnXi2}), we obtain from \eqs (\ref{ns,r}) that $n_s=0.961$ and $r=7.7\cdot 10^{-11}$. Both values are consistent with recent measurements of the cosmic microwave background \cite{Akrami:2018odb}. 

\section{Conclusions} 
We have considered the Standard Model with a conformally-invariant Higgs potential and proposed a model for how it can be merged with General Relativity. The two key ingredients are the non-minimal coupling of the Higgs field to the Ricci scalar and the Palatini formulation of gravity. No new degrees of freedom are introduced beyond those of the SM and the graviton. We have shown that after regulating the theory with an exemplary higher-dimensional operator, electroweak symmetry breaking can take place due to a singular gravitational-scalar instanton. In this way, an exponential suppression of the weak scale as compared to the Planck mass is naturally achieved. Moreover, such a setup offers the possibility to calculate the value of the former. Finally, the same theory leads to successful inflation with the Higgs boson as inflaton. 
Since the scale of violation of tree-level unitarity lies above inflationary energies, the Higgs potential during inflation can be determined from the low-energy parameters of the Standard Model, provided that corrections due to the strong-coupling regime at higher scales are not unnaturally large. 
This makes it possible to test inflationary physics at collider experiments and vice versa.

\begin{acknowledgments}
 We thank Fedor Bezrukov, Georgios Karananas, Marco Michel and Javier Rubio for useful discussions and comments. This work was supported by ERC-AdG-2015 grant 694896 and by the Swiss National Science Foundation Excellence grant 200020B\underline{ }182864.\\
\end{acknowledgments}

\section*{Appendix A: RG running of the Higgs mass.} 
The goal of this appendix is to study the influence of the non-minimal coupling $\xi$ on the running of the Higgs mass. Unlike in the rest of our work, we will first be more general and allow for a non-zero Higgs mass $m_H$, \ie we add to Lagrangian \eqref{jordanFrame} the term $-m_H^2/2\, h^2$. In the Einstein frame, this leads to
\begin{equation}
\begin{split}
\delta \mathcal{L}& = -\frac{m_H^2 M_P^2}{2\xi} \left(\frac{\tanh\left(\frac{\sqrt{\xi}\chi}{M_P}\right)}{ \cosh\left(\frac{\sqrt{\xi}\chi}{M_P}\right)}\right)^2\\
&\approx -\frac{1}{2} m_H^2 \chi^2 + \frac{5 m_H^2 \xi}{6 M_P^2}\chi^4 \,,
\end{split}
\end{equation}
where we expanded up to $4^{\text{th}}$ order in $\chi$ in the second step. Higher-order terms are suppressed by powers of $M_P/\sqrt{\xi}$ and therefore are subleading. Since all kinetic terms are canonical in the Einstein frame, we can apply the standard RG equations of a massive self-interacting scalar field. We conclude that the leading contribution of the non-minimal coupling to the $\beta_m$-function of the Higgs mass is
\begin{equation}
\delta \beta_m = \frac{-5 m_H^2 \xi}{4 \pi^2 M_P^2} m_H^2 \,.
\end{equation}
The novelty of this term consists in the fact that it is no longer suppressed by any coupling constant. However, it still vanishes for $m_H\rightarrow 0$. Thus, the Higgs mass is predictable in the theory \eqref{theory}.

\section*{Appendix B: Scale of unitarity violation in Palatini scenario.} 
In the metric scenario, we know that the scale of unitarity violation increases in a non-trivial background \cite{Bezrukov:2010jz}. We want to investigate if the same happens in the Palatini case. To this end, we follow the analysis of \cite{Bezrukov:2010jz} and study the scattering of gauge bosons. In unitary gauge, the Higgs boson interacts with a gauge boson $A_\mu$ via the term
\begin{equation} \label{modifiedGaugeBoson}
g'^{2} \frac{h^2}{\Omega^2} (A_\mu)^2 \,,
\end{equation}
where $g'$ is the weak coupling constant. We observe that at high energies, the coupling of the Higgs with gauge bosons is weaker than without non-minimal coupling. Therefore, the growth of the amplitudes involving longitudinal gauge bosons can no longer be compensated by scattering with Higgs particles. The compensation starts to fail as soon as $\Omega$ deviates from $1$, \ie when $h\gtrsim M_P/\sqrt{\xi}$ (equivalently $\chi\gtrsim M_P/\sqrt{\xi}$). At this point, the amplitudes $\mathcal{M}$ of the longitudinal gauge bosons grow as $\mathcal{M}\sim E/m_a$, where $E$ is the characteristic energy of the process and $m_a$ is the mass of gauge bosons. Using that $m_a \approx g' M_P/\sqrt{\xi}$ for $h\gtrsim M_P/\sqrt{\xi}$, we obtain the amplitude
\begin{equation}
\mathcal{M} \sim \frac{\sqrt{\xi} E}{g' M_P}\,.
\end{equation}
Up to the factor $g'$, which is of order $1$, the scale of unitarity violation therefore remains at $M_P/\sqrt{\xi}$ even at large background field values. We emphasize, however, that a more detailed study of the background-dependence of the scale of unitarity violation remains to be done.

\section*{Appendix C: RG running of the non-minimal coupling $\xi$.} 
In our analysis, we have neglected the RG running of the non-minimal coupling $\xi$. We want to study if this approximation is justified. Since at high energies the Higgs mass scales as $M_P/\sqrt{\xi}$ both in the metric and Palatini scenario, $\xi$ obeys the same RG equation in both cases, namely \cite{Bezrukov:2009db, Bezrukov:2017dyv}
\begin{equation} \label{betaXi}
\mu\, \partial_\mu \xi = -\frac{\alpha}{16 \pi^2}\xi \,,
\end{equation}
where $\alpha = 3/2 g'^{2} + 3 g^2 - 6 y_t^2$. Here $g'$ and $g$ are the $SU(2)$ and $U(1)$ gauge couplings of the Standard Model, respectively. In the approximation of constant $\alpha$, \eq \eqref{betaXi} is solved by
\begin{equation}
\xi(\mu) = \xi_0 \left(\frac{\mu_0}{\mu}\right)^{\frac{\alpha}{16 \pi^2}} \,,
\end{equation}
where $\mu_0$ represents a reference energy scale. 

As derived in \cite{SecondPaper}, the high-energy value of the top Yukawa coupling is $y_t\approx0.43$. Using the same method, we deduce that $g'\approx 0.44$ and $g\approx 0.53$.\footnote
{We thank Fedor Bezrukov for kindly providing us with a script to do so.}
It turns out that the contributions of the different couplings largely cancel, $\alpha \approx 0.03$. Both for the computation of the gravitational instanton and inflation, only field values $\chi \gtrsim M_P/\sqrt{\xi}$ are relevant. In this region of parameter space, the renormalization scale is bounded as $\mu \gtrsim y_t M_P/\sqrt{\xi}\  \tanh(1)$. Thus, the non-minimal coupling can at most vary by $\Delta \xi/\xi \approx 1-\tanh(1)^{\frac{\alpha}{16 \pi^2}}\approx 5 \cdot 10^{-5}$. Nevertheless, one can wonder if such a change of $\xi$, albeit small, can affect the flatness of the inflationary potential. Numerical analysis analogous to the one performed \cite{SecondPaper} shows that this is not the case, \ie all predictions remain invariant. It is interesting to note, however, that for $\alpha \sim 1$ small values of $\xi$ near the lower bound \eqref{BoundOnXi2} are no longer viable.

\section*{Appendix D: path integral measure.} 
As explained in the main text, we neglect any non-trivial factors in the measure of the path integral in \eq \eqref{PathIntegral2}. The reason is that they do not change the leading-order result of the saddle point-solution. In order to go beyond the leading order, one would have to compute all subleading contributions. They arise not only from the measure factors, but \eg also from the integration over the fields $\chi$ and $\hat{g}_{\mu\nu}$.
For completeness, and in order to illustrate these points, here we discuss the different contributions to the measure.

First, we do not take into account the higher-dimensional operator \eqref{OpDelta} and only consider a fundamental scalar field $h$ with the Lagrangian
\begin{equation}
\mathcal{L} = \frac{1}{2 \Omega^2} \left(\partial_\mu h\right)^2 - V \,.
\end{equation}
This corresponds to the Higgs part of the action \eqref{jordanFrame}, after performing the conformal transformation \eqref{Omega}. For the present discussion, $\Omega$ and $V$ could be arbitrary functions of $h$ and other fields, but they must be independent of $\partial_\mu h$. The corresponding conjugate momentum is 
\begin{equation}
\Pi = 1/\Omega^2\, \partial_0 h \,.
\end{equation}
Consequently, the Hamiltonian density reads
\begin{equation}
\mathcal{H} = \frac{\Omega^2}{2} \left(\Pi^2 + (\vec{\partial}h)^2\right) + V \,.
\end{equation}

In terms of field variable and its conjugate, the path integral (in Lorentzian signature) is
\begin{subequations} 
	\begin{align}
	\mathcal{P} & = \left(\prod_x \int \diff h(x) \diff \Pi(x)\right) \exp\left\{i \int \ldiff[4]{x} \Pi \partial_0 h - \mathcal{H}\right\}\\
	&= \left(\prod_x \int \diff h(x)\right)  \exp\left\{i \int \ldiff[4]{x} \mathcal{L}\right\}\\ & \cdot  \left(\prod_x  \diff \Pi(x) \exp\left\{-i \frac{\Omega^2}{2} \left(\Pi - \frac{1}{\Omega^2} \partial_0 h \right)^2 \right\} \right) \,.
	\end{align}
\end{subequations}
Performing the Gaussian integral in the last line, we obtain, up to constant factors:
\begin{equation} \label{modifiedMeasureH}
\mathcal{P} = \left(\prod_x \int \frac{\diff h(x)}{\Omega}\right)  \exp\left\{i \int \ldiff[4]{x} \mathcal{L}\right\} \,.
\end{equation}
In this way, we have derived the path integral measure of $h$ in the presence of a non-canonical kinetic term.\footnote
{The measure in \eq \eqref{modifiedMeasureH} is scale-invariant for $h \rightarrow \infty$. That a theory of Higgs inflation should obey this property has already been proposed long ago \cite{Bezrukov_2009} (this corresponds to ``prescription I'' in the language of \cite{Bezrukov_2009}). Further discussion of the choice of path integral measure can be found in \cite{Falls:2018olk}.}
Finally, we perform the transformation to the canonical field $\chi$. Since $\diff h = \Omega \diff \chi$ according to \eq \eqref{fieldTransformation}, we obtain
\begin{equation}
\mathcal{P} = \left(\prod_x \int \diff \chi(x)\right)  \exp\left\{i \int \ldiff[4]{x} \mathcal{L}\right\} \,.
\end{equation}
Thus, there is no additional contribution to the measure. Any field with a canonical kinetic term should exhibit this property \cite{Fradkin:1974df}.

Next we take into account the higher-order operator \eqref{OpDelta}. The conjugate momentum becomes 
\begin{equation} \label{modifiedCanonicalMomentum}
\Pi = \frac{1}{\Omega^2} \partial_0 h \left(1+ \frac{6 \delta}{M_P^8 \Omega^6} \left(1 + \frac{\delta}{\Omega^2}\right) (\partial_\mu h)^4 \right) \,.
\end{equation}
Consequently, the integral over $\Pi$ is no longer Gaussian and we cannot perform it explicitly. However, we can evaluate it in the saddle point approximation (see \eg \cite{Henneaux:1992ig}). It turns out that to leading order, the stationary point of $\Pi$ is still given by \eq \eqref{modifiedCanonicalMomentum}. Repeating the same steps as above, we obtain
\begin{subequations} \label{singularMeasureFactor}
	\begin{align}
	\mathcal{P} & \approx \left(\prod_x \int \diff \chi(x)\, m(x)\right)  \exp\left\{i \int \ldiff[4]{x} \mathcal{L}\right\}\\
	& = \left(\prod_x \int \diff \chi(x) \right)  \exp\left\{i \int \ldiff[4]{x}\left(\mathcal{L} -i \delta^{(4)}(0)\ln m(x)\right)\right\} \,,
	\end{align}
\end{subequations}
where
\begin{equation}
m(x) = \sqrt{1+ \frac{6 \delta}{M_P^8 \Omega^6} \left(1 + \frac{\delta}{\Omega^2}\right) \big((\partial_\mu h)^4 + 4 (\partial_\mu h)^2 (\partial_0 h)^2\big)}\,.
\end{equation}
We see that the additional contribution to the measure is singular. Such factors have long been known \cite{Lee:1962vm}, and they cancel in a consistent theory (see \eg also \cite{Salam:1971sp, Gerstein:1971fm}). This makes it evident that only after a complete study of fluctuations around the saddle-point solution, a finite correction can be obtained.

Finally, we briefly refer to results that are relevant for the gravitational part of the path integral. First, the integration over the connection in Palatini gravity leads to the same measure as in the metric theory \cite{Aros:2003bi}. However, the measure in metric gravity is non-trivial. It leads to the same singular factor as displayed in \eq \eqref{singularMeasureFactor}, where now (see \eg \cite{Aros:2003bi})
\begin{equation}
	m(x) = (\det G_{\alpha\beta\gamma\delta})^{-1/2} \;,
\end{equation}
with
\begin{equation}\label{G}
	G^{\alpha\beta\gamma\delta}(g)=\sqrt{g}\left( g^{\alpha\gamma}g^{\beta\delta}+g^{\alpha\delta}g^{\beta\gamma}-2g^{\alpha\beta}g^{\gamma\delta} \right) \;.
\end{equation}
As in the case of the scalar field, the non-trivial contribution of the measure only becomes meaningful after taking into account other corrections to the saddle-point solution.

\section*{Appendix E: instanton solution.} 
\begin{figure*}[t]
	\begin{center}
		\begin{minipage}{0.49\linewidth}
			\center{\includegraphics[scale=0.65]{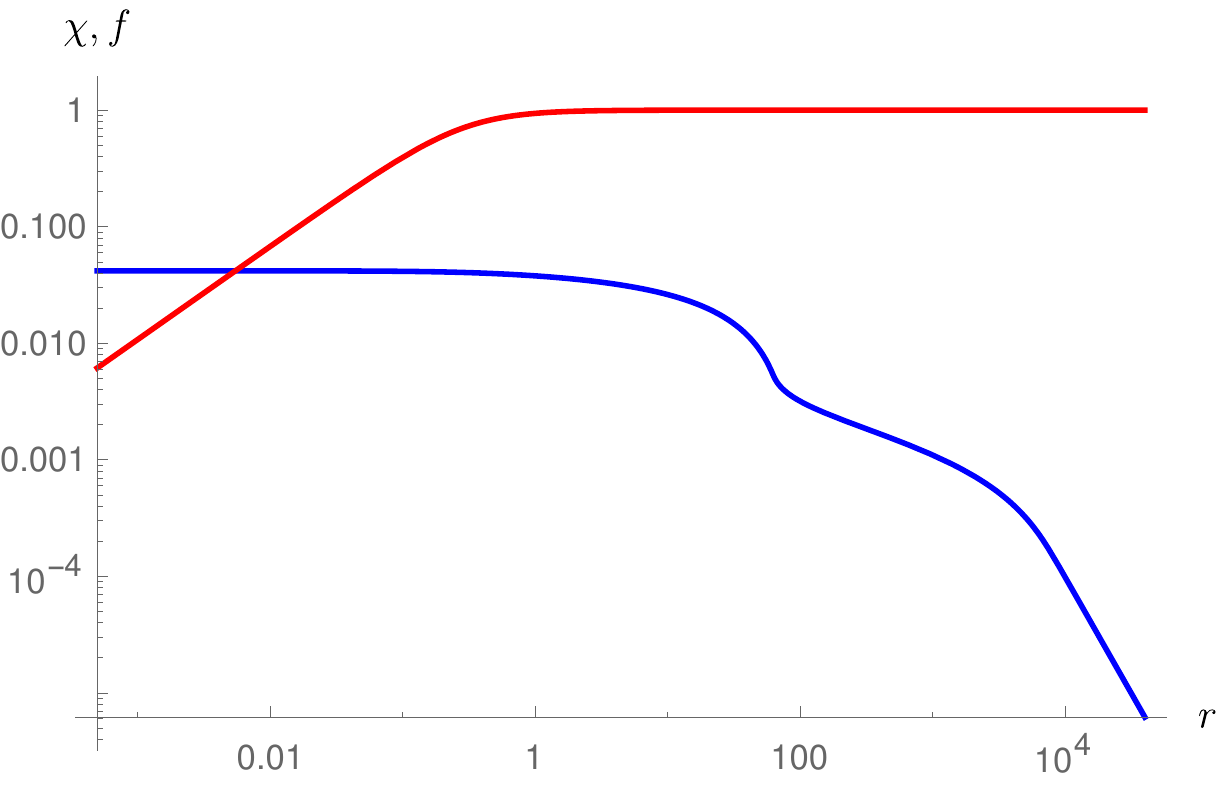}} \\(a)
		\end{minipage}
		\begin{minipage}{0.49\linewidth}
			\center{\includegraphics[scale=0.65]{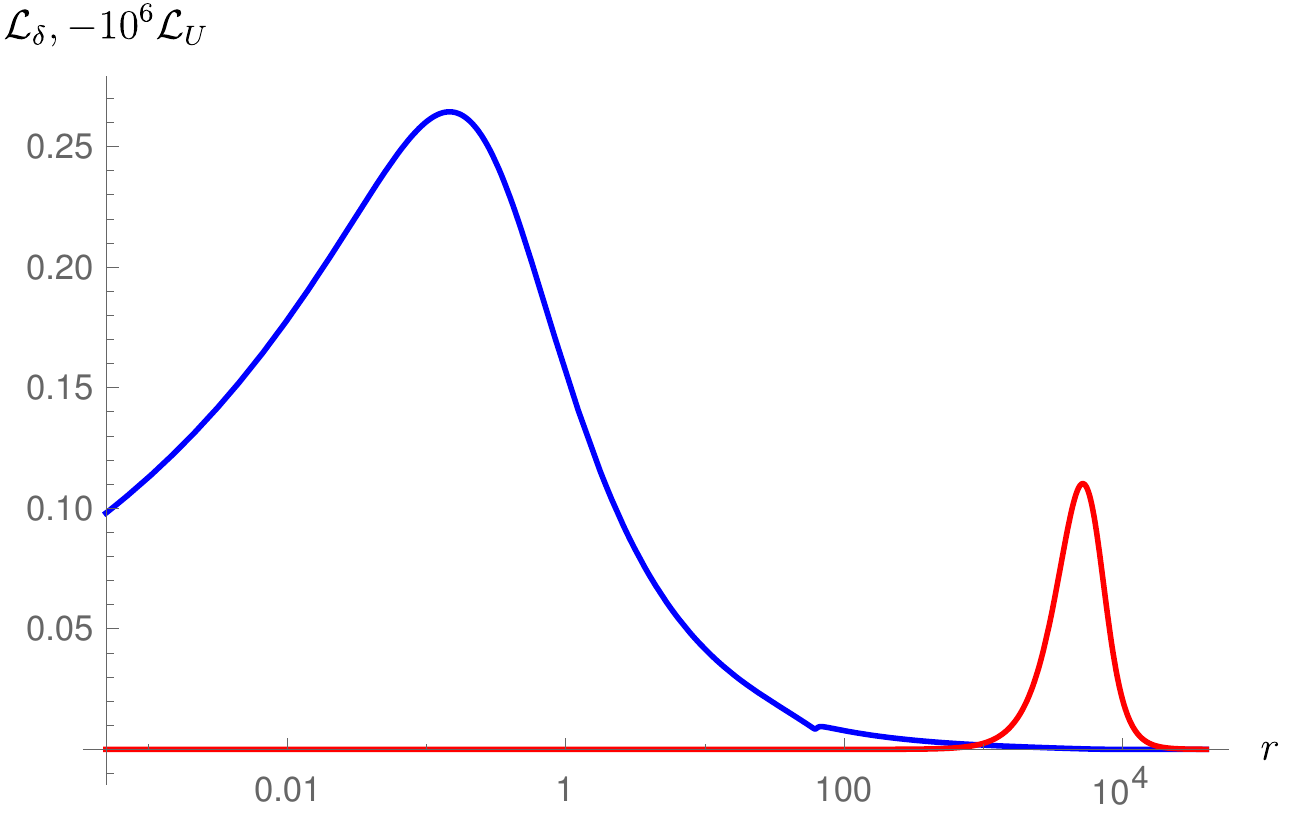}} \\(b)
		\end{minipage}
		\caption{(a) The field $\chi$ (blue) and the metric function $f$ (red) of the instanton solution corresponding to $\xi=10^7$, $\delta=0.887\xi^2$ (chosen so that $B=B_0$), and $\lambda=10^{-3}$. (b) Contributions $\mathcal{L}_\delta$ (blue) and $\mathcal{L}_U$ (red, scaled up $10^6$ times) to the Lagrangian for the same solution. All dimensionful quantities are in the units of $M_P$. }
		\label{fig:plot}
	\end{center}
\end{figure*}

In order to solve the full equations of motion \eqref{EoMs} numerically, we implement a shooting method. It relies on the fact that $f(r) = 1$ and $\chi(r)\sim r^{-2}$ for large $r$. Therefore, we can choose a sufficiently large value of $r$ and then demand that $f(r)=1$ and $r^2\chi(r) = c$ at this point, where $c$ initially is an arbitrary value. Using these boundary conditions, we solve the coupled differential equations \eqref{EoMs}. Subsequently, we check if the boundary condition at the origin, \ie \eq (\ref{EOMSD3}), is fulfilled. We change $c$ accordingly until the discrepancy between the \lhs and \rhs of \eq (\ref{EOMSD3}) is tolerably small.
The left panel of \fig \ref{fig:plot} shows the solutions $f(r)$ and $\chi(r)$ for exemplary values of the parameters of the theory.

The full instanton action is given by
\begin{equation}\label{FullAction}
B=-\sqrt{\xi}\chi(0)+2\pi^2\int_0^\infty dr\:(\mathcal{L}_\delta+\mathcal{L}_U) \;,
\end{equation}
with
\begin{equation}
\mathcal{L}_\delta=\frac{2\delta r^3G(\chi)\chi'^6}{f^5} \;, ~~~ \mathcal{L}_U=-f r^3U(\chi) \;.
\end{equation}
The right panel of \fig \ref{fig:plot} shows $\mathcal{L}_\delta$ and $\mathcal{L}_U$ as functions of $r$ for the exemplary solution. We see that the contribution to the action from the potential term is negligible. The total value of $B$ results from the balance between the negative source term and the positive higher-dimensional term.

Let us now discuss the validity of the semiclassical approximation in eq.~\eqref{Hierarchy} with the above instanton solution. We would like to use a scaling argument similar to eq.~\eqref{scaling} to show that a large parameter can always be extracted from the action \eqref{FullAction}. As before, we make use of the spherical symmetry of the solution and of the fact that $f\to 0$ as $r \to 0$. The analysis is complicated by the presence of the higher-dimensional term \eqref{OpDelta}. We will consider separately the inner region of the instanton, where $G(\chi)\approx 1$, and the outer region, where $G(\chi)\gg 1$. From the right panel of fig.~\ref{fig:plot} we see that both regions contribute to the instanton action. From the left panel of fig.~\ref{fig:plot} it is evident that in the inner region $f\ll 1$ while in the outer region $f\approx 1$.

Our goal is to factor out the leading-order $\xi$-dependence of the action $B$, as in eq.~\eqref{DimLessAction}. For a given value of $\xi$, we choose $\delta$ such that $B=\ln(M_P/(\sqrt{\xi}M_F))$ (see fig.~\ref{fig:XiVsDelta}). Numerical analysis shows that $\delta\sim\xi^\sigma$ where $\sigma\approx 2.24$. Our argument, however, does not depend on a particular value of $\sigma$, and we shall keep it arbitrary. Next, we make the following rescaling of the variables entering the action:
\begin{equation}
r\to \alpha\, r/M_P \;, ~~~ \chi\to \beta\, M_P\chi \;, ~~~ f\to \gamma\, f \;.
\end{equation}
To fix $\alpha$, $\beta$ and $\gamma$ as functions of $\xi$, we study the contributions to $B$ from the source term $B_\chi$, the curvature term $B_R$ and the higher-dimensional operator $B_\delta$. Consider first the inner region of the instanton, $G(\chi)\approx 1$. Then, the three contributions scale as
\begin{equation}
B_\chi\sim\xi^{1/2} \beta \;, ~~~ B_R\sim\frac{\alpha^2}{\gamma} \;, ~~~ B_\delta\sim \frac{\xi^{\sigma} \beta^6}{\alpha^2 \gamma^5} \;.
\end{equation}
Now we demand that all three terms exhibit the same scaling, $B_\chi \sim B_R \sim B_\delta\sim S$. Here $S$ is a function of $\xi$, which is fixed by the solution. We shall also keep it arbitrary since the argument is insensitive to the choice of $S$. Then, we obtain
\begin{equation}\label{Scaling1}
\alpha=S^{5/6}\, \xi^{(\sigma-3)/12}\;, ~ \beta= S\, \xi^{-1/2} \;, ~ \gamma= S^{2/3} \, \xi^{(\sigma-3)/6} \;.
\end{equation}
Consider now the outer region of the instanton, $G(\chi)\gg 1$. In this case $f\approx 1$, hence the only contribution to the action comes from the higher-dimensional term. It scales as
\begin{equation}
B_\delta\sim\frac{\xi^{2\sigma}\beta^6}{\alpha^2} \;.
\end{equation}
Demanding that this contribution also yields $S$ leads to
\begin{equation}\label{Scaling2}
\alpha= S^{-1/2} \, \xi^{(2\sigma-3)/2} \;, ~~~ \beta= \xi^{-\frac{1}{2}} \;,
\end{equation}
where $\beta$ is fixed by the requirement that the argument of cosh in $G(\chi)$ no longer depends on $\xi$. With the scalings \eqref{Scaling1} and \eqref{Scaling2} we have achieved our goal of extracting all large numbers from the integrals in the action.

\bibliography{Refs}

\begin{thebibliography}{77}%
\makeatletter
\providecommand \@ifxundefined [1]{%
 \@ifx{#1\undefined}
}%
\providecommand \@ifnum [1]{%
 \ifnum #1\expandafter \@firstoftwo
 \else \expandafter \@secondoftwo
 \fi
}%
\providecommand \@ifx [1]{%
 \ifx #1\expandafter \@firstoftwo
 \else \expandafter \@secondoftwo
 \fi
}%
\providecommand \natexlab [1]{#1}%
\providecommand \enquote  [1]{``#1''}%
\providecommand \bibnamefont  [1]{#1}%
\providecommand \bibfnamefont [1]{#1}%
\providecommand \citenamefont [1]{#1}%
\providecommand \href@noop [0]{\@secondoftwo}%
\providecommand \href [0]{\begingroup \@sanitize@url \@href}%
\providecommand \@href[1]{\@@startlink{#1}\@@href}%
\providecommand \@@href[1]{\endgroup#1\@@endlink}%
\providecommand \@sanitize@url [0]{\catcode `\\12\catcode `\$12\catcode
  `\&12\catcode `\#12\catcode `\^12\catcode `\_12\catcode `\%12\relax}%
\providecommand \@@startlink[1]{}%
\providecommand \@@endlink[0]{}%
\providecommand \url  [0]{\begingroup\@sanitize@url \@url }%
\providecommand \@url [1]{\endgroup\@href {#1}{\urlprefix }}%
\providecommand \urlprefix  [0]{URL }%
\providecommand \Eprint [0]{\href }%
\providecommand \doibase [0]{https://doi.org/}%
\providecommand \selectlanguage [0]{\@gobble}%
\providecommand \bibinfo  [0]{\@secondoftwo}%
\providecommand \bibfield  [0]{\@secondoftwo}%
\providecommand \translation [1]{[#1]}%
\providecommand \BibitemOpen [0]{}%
\providecommand \bibitemStop [0]{}%
\providecommand \bibitemNoStop [0]{.\EOS\space}%
\providecommand \EOS [0]{\spacefactor3000\relax}%
\providecommand \BibitemShut  [1]{\csname bibitem#1\endcsname}%
\let\auto@bib@innerbib\@empty
\bibitem [{\citenamefont {Aad}\ \emph {et~al.}(2012)\citenamefont {Aad} \emph
  {et~al.}}]{Aad:2012tfa}%
  \BibitemOpen
  \bibfield  {author} {\bibinfo {author} {\bibfnamefont {G.}~\bibnamefont
  {Aad}} \emph {et~al.} (\bibinfo {collaboration} {ATLAS}),\ }\bibfield
  {title} {\bibinfo {title} {{Observation of a new particle in the search for
  the Standard Model Higgs boson with the ATLAS detector at the LHC}},\ }\href
  {https://doi.org/10.1016/j.physletb.2012.08.020} {\bibfield  {journal}
  {\bibinfo  {journal} {Phys. Lett.}\ }\textbf {\bibinfo {volume} {B716}},\
  \bibinfo {pages} {1} (\bibinfo {year} {2012})},\ \Eprint
  {https://arxiv.org/abs/1207.7214} {arXiv:1207.7214 [hep-ex]} \BibitemShut
  {NoStop}%
\bibitem [{\citenamefont {Chatrchyan}\ \emph {et~al.}(2012)\citenamefont
  {Chatrchyan} \emph {et~al.}}]{Chatrchyan:2012xdj}%
  \BibitemOpen
  \bibfield  {author} {\bibinfo {author} {\bibfnamefont {S.}~\bibnamefont
  {Chatrchyan}} \emph {et~al.} (\bibinfo {collaboration} {CMS}),\ }\bibfield
  {title} {\bibinfo {title} {{Observation of a New Boson at a Mass of 125 GeV
  with the CMS Experiment at the LHC}},\ }\href
  {https://doi.org/10.1016/j.physletb.2012.08.021} {\bibfield  {journal}
  {\bibinfo  {journal} {Phys. Lett.}\ }\textbf {\bibinfo {volume} {B716}},\
  \bibinfo {pages} {30} (\bibinfo {year} {2012})},\ \Eprint
  {https://arxiv.org/abs/1207.7235} {arXiv:1207.7235 [hep-ex]} \BibitemShut
  {NoStop}%
\bibitem [{\citenamefont {Shaposhnikov}(2007)}]{Shaposhnikov:2007nj}%
  \BibitemOpen
  \bibfield  {author} {\bibinfo {author} {\bibfnamefont {M.}~\bibnamefont
  {Shaposhnikov}},\ }\bibfield  {title} {\bibinfo {title} {{Is there a new
  physics between electroweak and Planck scales?}},\ }in\ \href@noop {} {\emph
  {\bibinfo {booktitle} {{Astroparticle Physics: Current Issues, 2007 (APCI07)
  Budapest, Hungary, June 21-23, 2007}}}}\ (\bibinfo {year} {2007})\ \Eprint
  {https://arxiv.org/abs/0708.3550} {arXiv:0708.3550 [hep-th]} \BibitemShut
  {NoStop}%
\bibitem [{\citenamefont {Degrassi}\ \emph {et~al.}(2012)\citenamefont
  {Degrassi}, \citenamefont {Di~Vita}, \citenamefont {Elias-Miro},
  \citenamefont {Espinosa}, \citenamefont {Giudice}, \citenamefont {Isidori},\
  and\ \citenamefont {Strumia}}]{Degrassi:2012ry}%
  \BibitemOpen
  \bibfield  {author} {\bibinfo {author} {\bibfnamefont {G.}~\bibnamefont
  {Degrassi}}, \bibinfo {author} {\bibfnamefont {S.}~\bibnamefont {Di~Vita}},
  \bibinfo {author} {\bibfnamefont {J.}~\bibnamefont {Elias-Miro}}, \bibinfo
  {author} {\bibfnamefont {J.~R.}\ \bibnamefont {Espinosa}}, \bibinfo {author}
  {\bibfnamefont {G.~F.}\ \bibnamefont {Giudice}}, \bibinfo {author}
  {\bibfnamefont {G.}~\bibnamefont {Isidori}},\ and\ \bibinfo {author}
  {\bibfnamefont {A.}~\bibnamefont {Strumia}},\ }\bibfield  {title} {\bibinfo
  {title} {{Higgs mass and vacuum stability in the Standard Model at NNLO}},\
  }\href {https://doi.org/10.1007/JHEP08(2012)098} {\bibfield  {journal}
  {\bibinfo  {journal} {JHEP}\ }\textbf {\bibinfo {volume} {08}},\ \bibinfo
  {pages} {098}},\ \Eprint {https://arxiv.org/abs/1205.6497} {arXiv:1205.6497
  [hep-ph]} \BibitemShut {NoStop}%
\bibitem [{\citenamefont {Buttazzo}\ \emph {et~al.}(2013)\citenamefont
  {Buttazzo}, \citenamefont {Degrassi}, \citenamefont {Giardino}, \citenamefont
  {Giudice}, \citenamefont {Sala}, \citenamefont {Salvio},\ and\ \citenamefont
  {Strumia}}]{Buttazzo:2013uya}%
  \BibitemOpen
  \bibfield  {author} {\bibinfo {author} {\bibfnamefont {D.}~\bibnamefont
  {Buttazzo}}, \bibinfo {author} {\bibfnamefont {G.}~\bibnamefont {Degrassi}},
  \bibinfo {author} {\bibfnamefont {P.~P.}\ \bibnamefont {Giardino}}, \bibinfo
  {author} {\bibfnamefont {G.~F.}\ \bibnamefont {Giudice}}, \bibinfo {author}
  {\bibfnamefont {F.}~\bibnamefont {Sala}}, \bibinfo {author} {\bibfnamefont
  {A.}~\bibnamefont {Salvio}},\ and\ \bibinfo {author} {\bibfnamefont
  {A.}~\bibnamefont {Strumia}},\ }\bibfield  {title} {\bibinfo {title}
  {{Investigating the near-criticality of the Higgs boson}},\ }\href
  {https://doi.org/10.1007/JHEP12(2013)089} {\bibfield  {journal} {\bibinfo
  {journal} {JHEP}\ }\textbf {\bibinfo {volume} {12}},\ \bibinfo {pages}
  {089}},\ \Eprint {https://arxiv.org/abs/1307.3536} {arXiv:1307.3536 [hep-ph]}
  \BibitemShut {NoStop}%
\bibitem [{\citenamefont {Bezrukov}\ and\ \citenamefont
  {Shaposhnikov}(2015)}]{Bezrukov:2014ina}%
  \BibitemOpen
  \bibfield  {author} {\bibinfo {author} {\bibfnamefont {F.}~\bibnamefont
  {Bezrukov}}\ and\ \bibinfo {author} {\bibfnamefont {M.}~\bibnamefont
  {Shaposhnikov}},\ }\bibfield  {title} {\bibinfo {title} {{Why should we care
  about the top quark Yukawa coupling?}},\ }\href
  {https://doi.org/10.1134/S1063776115030152} {\bibfield  {journal} {\bibinfo
  {journal} {J. Exp. Theor. Phys.}\ }\textbf {\bibinfo {volume} {120}},\
  \bibinfo {pages} {335} (\bibinfo {year} {2015})},\ \bibinfo {note} {[Zh.
  Eksp. Teor. Fiz.147,389(2015)]},\ \Eprint {https://arxiv.org/abs/1411.1923}
  {arXiv:1411.1923 [hep-ph]} \BibitemShut {NoStop}%
\bibitem [{\citenamefont {Andreassen}\ \emph {et~al.}(2018)\citenamefont
  {Andreassen}, \citenamefont {Frost},\ and\ \citenamefont
  {Schwartz}}]{Andreassen:2017rzq}%
  \BibitemOpen
  \bibfield  {author} {\bibinfo {author} {\bibfnamefont {A.}~\bibnamefont
  {Andreassen}}, \bibinfo {author} {\bibfnamefont {W.}~\bibnamefont {Frost}},\
  and\ \bibinfo {author} {\bibfnamefont {M.~D.}\ \bibnamefont {Schwartz}},\
  }\bibfield  {title} {\bibinfo {title} {{Scale Invariant Instantons and the
  Complete Lifetime of the Standard Model}},\ }\href
  {https://doi.org/10.1103/PhysRevD.97.056006} {\bibfield  {journal} {\bibinfo
  {journal} {Phys. Rev.}\ }\textbf {\bibinfo {volume} {D97}},\ \bibinfo {pages}
  {056006} (\bibinfo {year} {2018})},\ \Eprint
  {https://arxiv.org/abs/1707.08124} {arXiv:1707.08124 [hep-ph]} \BibitemShut
  {NoStop}%
\bibitem [{\citenamefont {Asaka}\ \emph {et~al.}(2005)\citenamefont {Asaka},
  \citenamefont {Blanchet},\ and\ \citenamefont {Shaposhnikov}}]{Asaka:2005an}%
  \BibitemOpen
  \bibfield  {author} {\bibinfo {author} {\bibfnamefont {T.}~\bibnamefont
  {Asaka}}, \bibinfo {author} {\bibfnamefont {S.}~\bibnamefont {Blanchet}},\
  and\ \bibinfo {author} {\bibfnamefont {M.}~\bibnamefont {Shaposhnikov}},\
  }\bibfield  {title} {\bibinfo {title} {{The $\nu$MSM, dark matter and
  neutrino masses}},\ }\href {https://doi.org/10.1016/j.physletb.2005.09.070}
  {\bibfield  {journal} {\bibinfo  {journal} {Phys. Lett.}\ }\textbf {\bibinfo
  {volume} {B631}},\ \bibinfo {pages} {151} (\bibinfo {year} {2005})},\ \Eprint
  {https://arxiv.org/abs/hep-ph/0503065} {arXiv:hep-ph/0503065 [hep-ph]}
  \BibitemShut {NoStop}%
\bibitem [{\citenamefont {Asaka}\ and\ \citenamefont
  {Shaposhnikov}(2005)}]{Asaka:2005pn}%
  \BibitemOpen
  \bibfield  {author} {\bibinfo {author} {\bibfnamefont {T.}~\bibnamefont
  {Asaka}}\ and\ \bibinfo {author} {\bibfnamefont {M.}~\bibnamefont
  {Shaposhnikov}},\ }\bibfield  {title} {\bibinfo {title} {{The $\nu$MSM, dark
  matter and baryon asymmetry of the universe}},\ }\href
  {https://doi.org/10.1016/j.physletb.2005.06.020} {\bibfield  {journal}
  {\bibinfo  {journal} {Phys. Lett.}\ }\textbf {\bibinfo {volume} {B620}},\
  \bibinfo {pages} {17} (\bibinfo {year} {2005})},\ \Eprint
  {https://arxiv.org/abs/hep-ph/0505013} {arXiv:hep-ph/0505013 [hep-ph]}
  \BibitemShut {NoStop}%
\bibitem [{\citenamefont {Boyarsky}\ \emph {et~al.}(2009)\citenamefont
  {Boyarsky}, \citenamefont {Ruchayskiy},\ and\ \citenamefont
  {Shaposhnikov}}]{Boyarsky:2009ix}%
  \BibitemOpen
  \bibfield  {author} {\bibinfo {author} {\bibfnamefont {A.}~\bibnamefont
  {Boyarsky}}, \bibinfo {author} {\bibfnamefont {O.}~\bibnamefont
  {Ruchayskiy}},\ and\ \bibinfo {author} {\bibfnamefont {M.}~\bibnamefont
  {Shaposhnikov}},\ }\bibfield  {title} {\bibinfo {title} {{The Role of sterile
  neutrinos in cosmology and astrophysics}},\ }\href
  {https://doi.org/10.1146/annurev.nucl.010909.083654} {\bibfield  {journal}
  {\bibinfo  {journal} {Ann. Rev. Nucl. Part. Sci.}\ }\textbf {\bibinfo
  {volume} {59}},\ \bibinfo {pages} {191} (\bibinfo {year} {2009})},\ \Eprint
  {https://arxiv.org/abs/0901.0011} {arXiv:0901.0011 [hep-ph]} \BibitemShut
  {NoStop}%
\bibitem [{\citenamefont {Gildener}(1976)}]{Gildener:1976ai}%
  \BibitemOpen
  \bibfield  {author} {\bibinfo {author} {\bibfnamefont {E.}~\bibnamefont
  {Gildener}},\ }\bibfield  {title} {\bibinfo {title} {{Gauge Symmetry
  Hierarchies}},\ }\href {https://doi.org/10.1103/PhysRevD.14.1667} {\bibfield
  {journal} {\bibinfo  {journal} {Phys. Rev.}\ }\textbf {\bibinfo {volume}
  {D14}},\ \bibinfo {pages} {1667} (\bibinfo {year} {1976})}\BibitemShut
  {NoStop}%
\bibitem [{\citenamefont {Vissani}(1998)}]{Vissani:1997ys}%
  \BibitemOpen
  \bibfield  {author} {\bibinfo {author} {\bibfnamefont {F.}~\bibnamefont
  {Vissani}},\ }\bibfield  {title} {\bibinfo {title} {{Do experiments suggest a
  hierarchy problem?}},\ }\href {https://doi.org/10.1103/PhysRevD.57.7027}
  {\bibfield  {journal} {\bibinfo  {journal} {Phys. Rev.}\ }\textbf {\bibinfo
  {volume} {D57}},\ \bibinfo {pages} {7027} (\bibinfo {year} {1998})},\ \Eprint
  {https://arxiv.org/abs/hep-ph/9709409} {arXiv:hep-ph/9709409 [hep-ph]}
  \BibitemShut {NoStop}%
\bibitem [{\citenamefont {Farina}\ \emph {et~al.}(2013)\citenamefont {Farina},
  \citenamefont {Pappadopulo},\ and\ \citenamefont {Strumia}}]{Farina:2013mla}%
  \BibitemOpen
  \bibfield  {author} {\bibinfo {author} {\bibfnamefont {M.}~\bibnamefont
  {Farina}}, \bibinfo {author} {\bibfnamefont {D.}~\bibnamefont
  {Pappadopulo}},\ and\ \bibinfo {author} {\bibfnamefont {A.}~\bibnamefont
  {Strumia}},\ }\bibfield  {title} {\bibinfo {title} {{A modified naturalness
  principle and its experimental tests}},\ }\href
  {https://doi.org/10.1007/JHEP08(2013)022} {\bibfield  {journal} {\bibinfo
  {journal} {JHEP}\ }\textbf {\bibinfo {volume} {08}},\ \bibinfo {pages}
  {022}},\ \Eprint {https://arxiv.org/abs/1303.7244} {arXiv:1303.7244 [hep-ph]}
  \BibitemShut {NoStop}%
\bibitem [{\citenamefont {Palatini}(1919)}]{Palatini1919}%
  \BibitemOpen
  \bibfield  {author} {\bibinfo {author} {\bibfnamefont {A.}~\bibnamefont
  {Palatini}},\ }\bibfield  {title} {\bibinfo {title} {Deduzione invariantiva
  delle equazioni gravitazionali dal principio di hamilton},\ }\href
  {https://doi.org/10.1007/BF03014670} {\bibfield  {journal} {\bibinfo
  {journal} {Rendiconti del Circolo Matematico di Palermo}\ }\textbf {\bibinfo
  {volume} {43}},\ \bibinfo {pages} {203} (\bibinfo {year} {1919})}\BibitemShut
  {NoStop}%
\bibitem [{\citenamefont {Einstein}(1925)}]{Einstein1925}%
  \BibitemOpen
  \bibfield  {author} {\bibinfo {author} {\bibfnamefont {A.}~\bibnamefont
  {Einstein}},\ }\bibfield  {title} {\bibinfo {title} {{Einheitliche
  Feldtheorie von Gravitation und Elektrizit\"at}},\ }\href@noop {} {\bibfield
  {journal} {\bibinfo  {journal} {Sitzungsber. Preuss. Akad. Wiss}\ }\textbf
  {\bibinfo {volume} {414}} (\bibinfo {year} {1925})}\BibitemShut {NoStop}%
\bibitem [{\citenamefont {Coleman}\ and\ \citenamefont
  {Weinberg}(1973)}]{Coleman:1973jx}%
  \BibitemOpen
  \bibfield  {author} {\bibinfo {author} {\bibfnamefont {S.~R.}\ \bibnamefont
  {Coleman}}\ and\ \bibinfo {author} {\bibfnamefont {E.~J.}\ \bibnamefont
  {Weinberg}},\ }\bibfield  {title} {\bibinfo {title} {{Radiative Corrections
  as the Origin of Spontaneous Symmetry Breaking}},\ }\href
  {https://doi.org/10.1103/PhysRevD.7.1888} {\bibfield  {journal} {\bibinfo
  {journal} {Phys. Rev.}\ }\textbf {\bibinfo {volume} {D7}},\ \bibinfo {pages}
  {1888} (\bibinfo {year} {1973})}\BibitemShut {NoStop}%
\bibitem [{\citenamefont {Weinberg}(1976)}]{Weinberg:1976pe}%
  \BibitemOpen
  \bibfield  {author} {\bibinfo {author} {\bibfnamefont {S.}~\bibnamefont
  {Weinberg}},\ }\bibfield  {title} {\bibinfo {title} {{Mass of the Higgs
  Boson}},\ }\href {https://doi.org/10.1103/PhysRevLett.36.294} {\bibfield
  {journal} {\bibinfo  {journal} {Phys. Rev. Lett.}\ }\textbf {\bibinfo
  {volume} {36}},\ \bibinfo {pages} {294} (\bibinfo {year} {1976})}\BibitemShut
  {NoStop}%
\bibitem [{\citenamefont {Linde}(1977)}]{Linde:1977mm}%
  \BibitemOpen
  \bibfield  {author} {\bibinfo {author} {\bibfnamefont {A.~D.}\ \bibnamefont
  {Linde}},\ }\bibfield  {title} {\bibinfo {title} {{On the Vacuum Instability
  and the Higgs Meson Mass}},\ }\href
  {https://doi.org/10.1016/0370-2693(77)90664-5} {\bibfield  {journal}
  {\bibinfo  {journal} {Phys. Lett.}\ }\textbf {\bibinfo {volume} {70B}},\
  \bibinfo {pages} {306} (\bibinfo {year} {1977})}\BibitemShut {NoStop}%
\bibitem [{\citenamefont {'t~Hooft}(1973)}]{tHooft:1973mfk}%
  \BibitemOpen
  \bibfield  {author} {\bibinfo {author} {\bibfnamefont {G.}~\bibnamefont
  {'t~Hooft}},\ }\bibfield  {title} {\bibinfo {title} {{Dimensional
  regularization and the renormalization group}},\ }\href
  {https://doi.org/10.1016/0550-3213(73)90376-3} {\bibfield  {journal}
  {\bibinfo  {journal} {Nucl. Phys.}\ }\textbf {\bibinfo {volume} {B61}},\
  \bibinfo {pages} {455} (\bibinfo {year} {1973})}\BibitemShut {NoStop}%
\bibitem [{\citenamefont {'t~Hooft}\ and\ \citenamefont
  {Veltman}(1974)}]{tHooft:1974toh}%
  \BibitemOpen
  \bibfield  {author} {\bibinfo {author} {\bibfnamefont {G.}~\bibnamefont
  {'t~Hooft}}\ and\ \bibinfo {author} {\bibfnamefont {M.~J.~G.}\ \bibnamefont
  {Veltman}},\ }\bibfield  {title} {\bibinfo {title} {{One loop divergencies in
  the theory of gravitation}},\ }\href@noop {} {\bibfield  {journal} {\bibinfo
  {journal} {Ann. Inst. H. Poincare Phys. Theor.}\ }\textbf {\bibinfo {volume}
  {A20}},\ \bibinfo {pages} {69} (\bibinfo {year} {1974})}\BibitemShut
  {NoStop}%
\bibitem [{\citenamefont {Witten}(1981)}]{Witten:1980ez}%
  \BibitemOpen
  \bibfield  {author} {\bibinfo {author} {\bibfnamefont {E.}~\bibnamefont
  {Witten}},\ }\bibfield  {title} {\bibinfo {title} {{Cosmological Consequences
  of a Light Higgs Boson}},\ }\href
  {https://doi.org/10.1016/0550-3213(81)90182-6} {\bibfield  {journal}
  {\bibinfo  {journal} {Nucl. Phys.}\ }\textbf {\bibinfo {volume} {B177}},\
  \bibinfo {pages} {477} (\bibinfo {year} {1981})}\BibitemShut {NoStop}%
\bibitem [{\citenamefont {Froggatt}\ and\ \citenamefont
  {Nielsen}(1996)}]{Froggatt:1995rt}%
  \BibitemOpen
  \bibfield  {author} {\bibinfo {author} {\bibfnamefont {C.~D.}\ \bibnamefont
  {Froggatt}}\ and\ \bibinfo {author} {\bibfnamefont {H.~B.}\ \bibnamefont
  {Nielsen}},\ }\bibfield  {title} {\bibinfo {title} {{Standard model
  criticality prediction: Top mass 173 +- 5-GeV and Higgs mass 135 +- 9-GeV}},\
  }\href {https://doi.org/10.1016/0370-2693(95)01480-2} {\bibfield  {journal}
  {\bibinfo  {journal} {Phys. Lett.}\ }\textbf {\bibinfo {volume} {B368}},\
  \bibinfo {pages} {96} (\bibinfo {year} {1996})},\ \Eprint
  {https://arxiv.org/abs/hep-ph/9511371} {arXiv:hep-ph/9511371 [hep-ph]}
  \BibitemShut {NoStop}%
\bibitem [{\citenamefont {Shaposhnikov}\ and\ \citenamefont
  {Shkerin}(2018{\natexlab{a}})}]{Shaposhnikov:2018xkv}%
  \BibitemOpen
  \bibfield  {author} {\bibinfo {author} {\bibfnamefont {M.}~\bibnamefont
  {Shaposhnikov}}\ and\ \bibinfo {author} {\bibfnamefont {A.}~\bibnamefont
  {Shkerin}},\ }\bibfield  {title} {\bibinfo {title} {{Conformal symmetry:
  towards the link between the Fermi and the Planck scales}},\ }\href
  {https://doi.org/10.1016/j.physletb.2018.06.068} {\bibfield  {journal}
  {\bibinfo  {journal} {Phys. Lett.}\ }\textbf {\bibinfo {volume} {B783}},\
  \bibinfo {pages} {253} (\bibinfo {year} {2018}{\natexlab{a}})},\ \Eprint
  {https://arxiv.org/abs/1803.08907} {arXiv:1803.08907 [hep-th]} \BibitemShut
  {NoStop}%
\bibitem [{\citenamefont {Bezrukov}\ and\ \citenamefont
  {Shaposhnikov}(2008)}]{Bezrukov:2007ep}%
  \BibitemOpen
  \bibfield  {author} {\bibinfo {author} {\bibfnamefont {F.~L.}\ \bibnamefont
  {Bezrukov}}\ and\ \bibinfo {author} {\bibfnamefont {M.}~\bibnamefont
  {Shaposhnikov}},\ }\bibfield  {title} {\bibinfo {title} {{The Standard Model
  Higgs boson as the inflaton}},\ }\href
  {https://doi.org/10.1016/j.physletb.2007.11.072} {\bibfield  {journal}
  {\bibinfo  {journal} {Phys. Lett.}\ }\textbf {\bibinfo {volume} {B659}},\
  \bibinfo {pages} {703} (\bibinfo {year} {2008})},\ \Eprint
  {https://arxiv.org/abs/0710.3755} {arXiv:0710.3755 [hep-th]} \BibitemShut
  {NoStop}%
\bibitem [{\citenamefont {Bauer}\ and\ \citenamefont
  {Demir}(2008)}]{Bauer:2008zj}%
  \BibitemOpen
  \bibfield  {author} {\bibinfo {author} {\bibfnamefont {F.}~\bibnamefont
  {Bauer}}\ and\ \bibinfo {author} {\bibfnamefont {D.~A.}\ \bibnamefont
  {Demir}},\ }\bibfield  {title} {\bibinfo {title} {{Inflation with Non-Minimal
  Coupling: Metric versus Palatini Formulations}},\ }\href
  {https://doi.org/10.1016/j.physletb.2008.06.014} {\bibfield  {journal}
  {\bibinfo  {journal} {Phys. Lett.}\ }\textbf {\bibinfo {volume} {B665}},\
  \bibinfo {pages} {222} (\bibinfo {year} {2008})},\ \Eprint
  {https://arxiv.org/abs/0803.2664} {arXiv:0803.2664 [hep-ph]} \BibitemShut
  {NoStop}%
\bibitem [{\citenamefont {Bezrukov}\ \emph {et~al.}(2011)\citenamefont
  {Bezrukov}, \citenamefont {Magnin}, \citenamefont {Shaposhnikov},\ and\
  \citenamefont {Sibiryakov}}]{Bezrukov:2010jz}%
  \BibitemOpen
  \bibfield  {author} {\bibinfo {author} {\bibfnamefont {F.}~\bibnamefont
  {Bezrukov}}, \bibinfo {author} {\bibfnamefont {A.}~\bibnamefont {Magnin}},
  \bibinfo {author} {\bibfnamefont {M.}~\bibnamefont {Shaposhnikov}},\ and\
  \bibinfo {author} {\bibfnamefont {S.}~\bibnamefont {Sibiryakov}},\ }\bibfield
   {title} {\bibinfo {title} {{Higgs inflation: consistency and
  generalisations}},\ }\href {https://doi.org/10.1007/JHEP01(2011)016}
  {\bibfield  {journal} {\bibinfo  {journal} {JHEP}\ }\textbf {\bibinfo
  {volume} {01}},\ \bibinfo {pages} {016}},\ \Eprint
  {https://arxiv.org/abs/1008.5157} {arXiv:1008.5157 [hep-ph]} \BibitemShut
  {NoStop}%
\bibitem [{\citenamefont {Escrivà}\ and\ \citenamefont
  {Germani}(2017)}]{Escriva:2016cwl}%
  \BibitemOpen
  \bibfield  {author} {\bibinfo {author} {\bibfnamefont {A.}~\bibnamefont
  {Escrivà}}\ and\ \bibinfo {author} {\bibfnamefont {C.}~\bibnamefont
  {Germani}},\ }\bibfield  {title} {\bibinfo {title} {{Beyond dimensional
  analysis: Higgs and new Higgs inflations do not violate unitarity}},\ }\href
  {https://doi.org/10.1103/PhysRevD.95.123526} {\bibfield  {journal} {\bibinfo
  {journal} {Phys. Rev. D}\ }\textbf {\bibinfo {volume} {95}},\ \bibinfo
  {pages} {123526} (\bibinfo {year} {2017})},\ \Eprint
  {https://arxiv.org/abs/1612.06253} {arXiv:1612.06253 [hep-ph]} \BibitemShut
  {NoStop}%
\bibitem [{\citenamefont {Fumagalli}\ \emph {et~al.}(2018)\citenamefont
  {Fumagalli}, \citenamefont {Mooij},\ and\ \citenamefont
  {Postma}}]{Fumagalli:2017cdo}%
  \BibitemOpen
  \bibfield  {author} {\bibinfo {author} {\bibfnamefont {J.}~\bibnamefont
  {Fumagalli}}, \bibinfo {author} {\bibfnamefont {S.}~\bibnamefont {Mooij}},\
  and\ \bibinfo {author} {\bibfnamefont {M.}~\bibnamefont {Postma}},\
  }\bibfield  {title} {\bibinfo {title} {{Unitarity and predictiveness in new
  Higgs inflation}},\ }\href {https://doi.org/10.1007/JHEP03(2018)038}
  {\bibfield  {journal} {\bibinfo  {journal} {JHEP}\ }\textbf {\bibinfo
  {volume} {03}},\ \bibinfo {pages} {038}},\ \Eprint
  {https://arxiv.org/abs/1711.08761} {arXiv:1711.08761 [hep-ph]} \BibitemShut
  {NoStop}%
\bibitem [{\citenamefont {Barbon}\ and\ \citenamefont
  {Espinosa}(2009)}]{Barbon:2009ya}%
  \BibitemOpen
  \bibfield  {author} {\bibinfo {author} {\bibfnamefont {J.~L.~F.}\
  \bibnamefont {Barbon}}\ and\ \bibinfo {author} {\bibfnamefont {J.~R.}\
  \bibnamefont {Espinosa}},\ }\bibfield  {title} {\bibinfo {title} {{On the
  Naturalness of Higgs Inflation}},\ }\href
  {https://doi.org/10.1103/PhysRevD.79.081302} {\bibfield  {journal} {\bibinfo
  {journal} {Phys. Rev.}\ }\textbf {\bibinfo {volume} {D79}},\ \bibinfo {pages}
  {081302} (\bibinfo {year} {2009})},\ \Eprint
  {https://arxiv.org/abs/0903.0355} {arXiv:0903.0355 [hep-ph]} \BibitemShut
  {NoStop}%
\bibitem [{\citenamefont {Burgess}\ \emph {et~al.}(2009)\citenamefont
  {Burgess}, \citenamefont {Lee},\ and\ \citenamefont
  {Trott}}]{Burgess:2009ea}%
  \BibitemOpen
  \bibfield  {author} {\bibinfo {author} {\bibfnamefont {C.~P.}\ \bibnamefont
  {Burgess}}, \bibinfo {author} {\bibfnamefont {H.~M.}\ \bibnamefont {Lee}},\
  and\ \bibinfo {author} {\bibfnamefont {M.}~\bibnamefont {Trott}},\ }\bibfield
   {title} {\bibinfo {title} {{Power-counting and the Validity of the Classical
  Approximation During Inflation}},\ }\href
  {https://doi.org/10.1088/1126-6708/2009/09/103} {\bibfield  {journal}
  {\bibinfo  {journal} {JHEP}\ }\textbf {\bibinfo {volume} {09}},\ \bibinfo
  {pages} {103}},\ \Eprint {https://arxiv.org/abs/0902.4465} {arXiv:0902.4465
  [hep-ph]} \BibitemShut {NoStop}%
\bibitem [{\citenamefont {Bezrukov}\ \emph {et~al.}(2015)\citenamefont
  {Bezrukov}, \citenamefont {Rubio},\ and\ \citenamefont
  {Shaposhnikov}}]{Bezrukov:2014ipa}%
  \BibitemOpen
  \bibfield  {author} {\bibinfo {author} {\bibfnamefont {F.}~\bibnamefont
  {Bezrukov}}, \bibinfo {author} {\bibfnamefont {J.}~\bibnamefont {Rubio}},\
  and\ \bibinfo {author} {\bibfnamefont {M.}~\bibnamefont {Shaposhnikov}},\
  }\bibfield  {title} {\bibinfo {title} {{Living beyond the edge: Higgs
  inflation and vacuum metastability}},\ }\href
  {https://doi.org/10.1103/PhysRevD.92.083512} {\bibfield  {journal} {\bibinfo
  {journal} {Phys. Rev.}\ }\textbf {\bibinfo {volume} {D92}},\ \bibinfo {pages}
  {083512} (\bibinfo {year} {2015})},\ \Eprint
  {https://arxiv.org/abs/1412.3811} {arXiv:1412.3811 [hep-ph]} \BibitemShut
  {NoStop}%
\bibitem [{\citenamefont {Ema}\ \emph {et~al.}(2017)\citenamefont {Ema},
  \citenamefont {Jinno}, \citenamefont {Mukaida},\ and\ \citenamefont
  {Nakayama}}]{Ema:2016dny}%
  \BibitemOpen
  \bibfield  {author} {\bibinfo {author} {\bibfnamefont {Y.}~\bibnamefont
  {Ema}}, \bibinfo {author} {\bibfnamefont {R.}~\bibnamefont {Jinno}}, \bibinfo
  {author} {\bibfnamefont {K.}~\bibnamefont {Mukaida}},\ and\ \bibinfo {author}
  {\bibfnamefont {K.}~\bibnamefont {Nakayama}},\ }\bibfield  {title} {\bibinfo
  {title} {{Violent Preheating in Inflation with Nonminimal Coupling}},\ }\href
  {https://doi.org/10.1088/1475-7516/2017/02/045} {\bibfield  {journal}
  {\bibinfo  {journal} {JCAP}\ }\textbf {\bibinfo {volume} {1702}}\bibfield
  {number} {\bibinfo  {number} { (02)},\ \bibinfo {pages} {045}},\ }\Eprint
  {https://arxiv.org/abs/1609.05209} {arXiv:1609.05209 [hep-ph]} \BibitemShut
  {NoStop}%
\bibitem [{\citenamefont {DeCross}\ \emph {et~al.}(2018)\citenamefont
  {DeCross}, \citenamefont {Kaiser}, \citenamefont {Prabhu}, \citenamefont
  {Prescod-Weinstein},\ and\ \citenamefont {Sfakianakis}}]{DeCross:2016cbs}%
  \BibitemOpen
  \bibfield  {author} {\bibinfo {author} {\bibfnamefont {M.~P.}\ \bibnamefont
  {DeCross}}, \bibinfo {author} {\bibfnamefont {D.~I.}\ \bibnamefont {Kaiser}},
  \bibinfo {author} {\bibfnamefont {A.}~\bibnamefont {Prabhu}}, \bibinfo
  {author} {\bibfnamefont {C.}~\bibnamefont {Prescod-Weinstein}},\ and\
  \bibinfo {author} {\bibfnamefont {E.~I.}\ \bibnamefont {Sfakianakis}},\
  }\bibfield  {title} {\bibinfo {title} {{Preheating after multifield inflation
  with nonminimal couplings, III: Dynamical spacetime results}},\ }\href
  {https://doi.org/10.1103/PhysRevD.97.023528} {\bibfield  {journal} {\bibinfo
  {journal} {Phys. Rev.}\ }\textbf {\bibinfo {volume} {D97}},\ \bibinfo {pages}
  {023528} (\bibinfo {year} {2018})},\ \Eprint
  {https://arxiv.org/abs/1610.08916} {arXiv:1610.08916 [astro-ph.CO]}
  \BibitemShut {NoStop}%
\bibitem [{\citenamefont {Bauer}\ and\ \citenamefont
  {Demir}(2011)}]{Bauer:2010jg}%
  \BibitemOpen
  \bibfield  {author} {\bibinfo {author} {\bibfnamefont {F.}~\bibnamefont
  {Bauer}}\ and\ \bibinfo {author} {\bibfnamefont {D.~A.}\ \bibnamefont
  {Demir}},\ }\bibfield  {title} {\bibinfo {title} {{Higgs-Palatini Inflation
  and Unitarity}},\ }\href {https://doi.org/10.1016/j.physletb.2011.03.042}
  {\bibfield  {journal} {\bibinfo  {journal} {Phys. Lett.}\ }\textbf {\bibinfo
  {volume} {B698}},\ \bibinfo {pages} {425} (\bibinfo {year} {2011})},\ \Eprint
  {https://arxiv.org/abs/1012.2900} {arXiv:1012.2900 [hep-ph]} \BibitemShut
  {NoStop}%
\bibitem [{\citenamefont {Shaposhnikov}\ \emph {et~al.}(2020)\citenamefont
  {Shaposhnikov}, \citenamefont {Shkerin},\ and\ \citenamefont
  {Zell}}]{SecondPaper}%
  \BibitemOpen
  \bibfield  {author} {\bibinfo {author} {\bibfnamefont {M.}~\bibnamefont
  {Shaposhnikov}}, \bibinfo {author} {\bibfnamefont {A.}~\bibnamefont
  {Shkerin}},\ and\ \bibinfo {author} {\bibfnamefont {S.}~\bibnamefont
  {Zell}},\ }\bibfield  {title} {\bibinfo {title} {{Quantum Effects in Palatini
  Higgs Inflation}},\ }\href {https://doi.org/10.1088/1475-7516/2020/07/064}
  {\bibfield  {journal} {\bibinfo  {journal} {JCAP}\ }\textbf {\bibinfo
  {volume} {07}},\ \bibinfo {pages} {064}},\ \Eprint
  {https://arxiv.org/abs/2002.07105} {arXiv:2002.07105 [hep-ph]} \BibitemShut
  {NoStop}%
\bibitem [{\citenamefont {Aydemir}\ \emph {et~al.}(2012)\citenamefont
  {Aydemir}, \citenamefont {Anber},\ and\ \citenamefont
  {Donoghue}}]{Aydemir:2012nz}%
  \BibitemOpen
  \bibfield  {author} {\bibinfo {author} {\bibfnamefont {U.}~\bibnamefont
  {Aydemir}}, \bibinfo {author} {\bibfnamefont {M.~M.}\ \bibnamefont {Anber}},\
  and\ \bibinfo {author} {\bibfnamefont {J.~F.}\ \bibnamefont {Donoghue}},\
  }\bibfield  {title} {\bibinfo {title} {{Self-healing of unitarity in
  effective field theories and the onset of new physics}},\ }\href
  {https://doi.org/10.1103/PhysRevD.86.014025} {\bibfield  {journal} {\bibinfo
  {journal} {Phys. Rev.}\ }\textbf {\bibinfo {volume} {D86}},\ \bibinfo {pages}
  {014025} (\bibinfo {year} {2012})},\ \Eprint
  {https://arxiv.org/abs/1203.5153} {arXiv:1203.5153 [hep-ph]} \BibitemShut
  {NoStop}%
\bibitem [{\citenamefont {Weinberg}(1979)}]{Weinberg:1980gg}%
  \BibitemOpen
  \bibfield  {author} {\bibinfo {author} {\bibfnamefont {S.}~\bibnamefont
  {Weinberg}},\ }\bibinfo {title} {{Ultraviolet divergences in quantum theories
  of gravitation}},\ in\ \href@noop {} {\emph {\bibinfo {booktitle} {{General
  Relativity}: {An Einstein Centenary Survey}}}}\ (\bibinfo {year} {1979})\
  pp.\ \bibinfo {pages} {790--831}\BibitemShut {NoStop}%
\bibitem [{\citenamefont {Reuter}(1998)}]{Reuter:1996cp}%
  \BibitemOpen
  \bibfield  {author} {\bibinfo {author} {\bibfnamefont {M.}~\bibnamefont
  {Reuter}},\ }\bibfield  {title} {\bibinfo {title} {{Nonperturbative evolution
  equation for quantum gravity}},\ }\href
  {https://doi.org/10.1103/PhysRevD.57.971} {\bibfield  {journal} {\bibinfo
  {journal} {Phys. Rev. D}\ }\textbf {\bibinfo {volume} {57}},\ \bibinfo
  {pages} {971} (\bibinfo {year} {1998})},\ \Eprint
  {https://arxiv.org/abs/hep-th/9605030} {arXiv:hep-th/9605030} \BibitemShut
  {NoStop}%
\bibitem [{\citenamefont {Dvali}\ and\ \citenamefont
  {Gomez}(2010)}]{Dvali:2010bf}%
  \BibitemOpen
  \bibfield  {author} {\bibinfo {author} {\bibfnamefont {G.}~\bibnamefont
  {Dvali}}\ and\ \bibinfo {author} {\bibfnamefont {C.}~\bibnamefont {Gomez}},\
  }\bibfield  {title} {\bibinfo {title} {{Self-Completeness of Einstein
  Gravity}},\ }\href@noop {} {\  (\bibinfo {year} {2010})},\ \Eprint
  {https://arxiv.org/abs/1005.3497} {arXiv:1005.3497 [hep-th]} \BibitemShut
  {NoStop}%
\bibitem [{\citenamefont {Dvali}\ \emph
  {et~al.}(2011{\natexlab{a}})\citenamefont {Dvali}, \citenamefont {Giudice},
  \citenamefont {Gomez},\ and\ \citenamefont {Kehagias}}]{Dvali:2010jz}%
  \BibitemOpen
  \bibfield  {author} {\bibinfo {author} {\bibfnamefont {G.}~\bibnamefont
  {Dvali}}, \bibinfo {author} {\bibfnamefont {G.~F.}\ \bibnamefont {Giudice}},
  \bibinfo {author} {\bibfnamefont {C.}~\bibnamefont {Gomez}},\ and\ \bibinfo
  {author} {\bibfnamefont {A.}~\bibnamefont {Kehagias}},\ }\bibfield  {title}
  {\bibinfo {title} {{UV-Completion by Classicalization}},\ }\href
  {https://doi.org/10.1007/JHEP08(2011)108} {\bibfield  {journal} {\bibinfo
  {journal} {JHEP}\ }\textbf {\bibinfo {volume} {08}},\ \bibinfo {pages}
  {108}},\ \Eprint {https://arxiv.org/abs/1010.1415} {arXiv:1010.1415 [hep-ph]}
  \BibitemShut {NoStop}%
\bibitem [{\citenamefont {Dvali}\ \emph
  {et~al.}(2011{\natexlab{b}})\citenamefont {Dvali}, \citenamefont {Gomez},\
  and\ \citenamefont {Kehagias}}]{Dvali:2011th}%
  \BibitemOpen
  \bibfield  {author} {\bibinfo {author} {\bibfnamefont {G.}~\bibnamefont
  {Dvali}}, \bibinfo {author} {\bibfnamefont {C.}~\bibnamefont {Gomez}},\ and\
  \bibinfo {author} {\bibfnamefont {A.}~\bibnamefont {Kehagias}},\ }\bibfield
  {title} {\bibinfo {title} {{Classicalization of Gravitons and Goldstones}},\
  }\href {https://doi.org/10.1007/JHEP11(2011)070} {\bibfield  {journal}
  {\bibinfo  {journal} {JHEP}\ }\textbf {\bibinfo {volume} {11}},\ \bibinfo
  {pages} {070}},\ \Eprint {https://arxiv.org/abs/1103.5963} {arXiv:1103.5963
  [hep-th]} \BibitemShut {NoStop}%
\bibitem [{\citenamefont {Shaposhnikov}\ and\ \citenamefont
  {Shkerin}(2018{\natexlab{b}})}]{Shaposhnikov:2018jag}%
  \BibitemOpen
  \bibfield  {author} {\bibinfo {author} {\bibfnamefont {M.}~\bibnamefont
  {Shaposhnikov}}\ and\ \bibinfo {author} {\bibfnamefont {A.}~\bibnamefont
  {Shkerin}},\ }\bibfield  {title} {\bibinfo {title} {{Gravity, Scale
  Invariance and the Hierarchy Problem}},\ }\href
  {https://doi.org/10.1007/JHEP10(2018)024} {\bibfield  {journal} {\bibinfo
  {journal} {JHEP}\ }\textbf {\bibinfo {volume} {10}},\ \bibinfo {pages}
  {024}},\ \Eprint {https://arxiv.org/abs/1804.06376} {arXiv:1804.06376
  [hep-th]} \BibitemShut {NoStop}%
\bibitem [{\citenamefont {Shkerin}(2019)}]{Shkerin:2019mmu}%
  \BibitemOpen
  \bibfield  {author} {\bibinfo {author} {\bibfnamefont {A.}~\bibnamefont
  {Shkerin}},\ }\bibfield  {title} {\bibinfo {title} {{Dilaton-assisted
  generation of the Fermi scale from the Planck scale}},\ }\href
  {https://doi.org/10.1103/PhysRevD.99.115018} {\bibfield  {journal} {\bibinfo
  {journal} {Phys. Rev.}\ }\textbf {\bibinfo {volume} {D99}},\ \bibinfo {pages}
  {115018} (\bibinfo {year} {2019})},\ \Eprint
  {https://arxiv.org/abs/1903.11317} {arXiv:1903.11317 [hep-th]} \BibitemShut
  {NoStop}%
\bibitem [{\citenamefont {Gibbons}(1977)}]{Gibbons:1977zz}%
  \BibitemOpen
  \bibfield  {author} {\bibinfo {author} {\bibfnamefont {G.~W.}\ \bibnamefont
  {Gibbons}},\ }\bibfield  {title} {\bibinfo {title} {{The Einstein Action of
  Riemannian Metrics and Its Relation to Quantum Gravity and Thermodynamics}},\
  }\href {https://doi.org/10.1016/0375-9601(77)90244-4} {\bibfield  {journal}
  {\bibinfo  {journal} {Phys. Lett.}\ }\textbf {\bibinfo {volume} {A61}},\
  \bibinfo {pages} {3} (\bibinfo {year} {1977})}\BibitemShut {NoStop}%
\bibitem [{\citenamefont {Gratton}\ and\ \citenamefont
  {Turok}(1999)}]{Gratton:1999ya}%
  \BibitemOpen
  \bibfield  {author} {\bibinfo {author} {\bibfnamefont {S.}~\bibnamefont
  {Gratton}}\ and\ \bibinfo {author} {\bibfnamefont {N.}~\bibnamefont
  {Turok}},\ }\bibfield  {title} {\bibinfo {title} {{Cosmological perturbations
  from the no boundary Euclidean path integral}},\ }\href
  {https://doi.org/10.1103/PhysRevD.60.123507} {\bibfield  {journal} {\bibinfo
  {journal} {Phys. Rev.}\ }\textbf {\bibinfo {volume} {D60}},\ \bibinfo {pages}
  {123507} (\bibinfo {year} {1999})},\ \Eprint
  {https://arxiv.org/abs/astro-ph/9902265} {arXiv:astro-ph/9902265 [astro-ph]}
  \BibitemShut {NoStop}%
\bibitem [{\citenamefont {Dadhich}\ and\ \citenamefont
  {Pons}(2012)}]{Dadhich:2010xa}%
  \BibitemOpen
  \bibfield  {author} {\bibinfo {author} {\bibfnamefont {N.}~\bibnamefont
  {Dadhich}}\ and\ \bibinfo {author} {\bibfnamefont {J.~M.}\ \bibnamefont
  {Pons}},\ }\bibfield  {title} {\bibinfo {title} {{On the equivalence of the
  Einstein-Hilbert and the Einstein-Palatini formulations of general relativity
  for an arbitrary connection}},\ }\href
  {https://doi.org/10.1007/s10714-012-1393-9} {\bibfield  {journal} {\bibinfo
  {journal} {Gen. Rel. Grav.}\ }\textbf {\bibinfo {volume} {44}},\ \bibinfo
  {pages} {2337} (\bibinfo {year} {2012})},\ \Eprint
  {https://arxiv.org/abs/1010.0869} {arXiv:1010.0869 [gr-qc]} \BibitemShut
  {NoStop}%
\bibitem [{\citenamefont {Polyakov}(1978)}]{Polyakov:1978vu}%
  \BibitemOpen
  \bibfield  {author} {\bibinfo {author} {\bibfnamefont {A.~M.}\ \bibnamefont
  {Polyakov}},\ }\bibfield  {title} {\bibinfo {title} {{Thermal Properties of
  Gauge Fields and Quark Liberation}},\ }\href
  {https://doi.org/10.1016/0370-2693(78)90737-2} {\bibfield  {journal}
  {\bibinfo  {journal} {Phys. Lett.}\ }\textbf {\bibinfo {volume} {72B}},\
  \bibinfo {pages} {477} (\bibinfo {year} {1978})}\BibitemShut {NoStop}%
\bibitem [{\citenamefont {Khlebnikov}\ \emph {et~al.}(1991)\citenamefont
  {Khlebnikov}, \citenamefont {Rubakov},\ and\ \citenamefont
  {Tinyakov}}]{Khlebnikov:1990ue}%
  \BibitemOpen
  \bibfield  {author} {\bibinfo {author} {\bibfnamefont {S.~{\relax Yu}.}\
  \bibnamefont {Khlebnikov}}, \bibinfo {author} {\bibfnamefont {V.~A.}\
  \bibnamefont {Rubakov}},\ and\ \bibinfo {author} {\bibfnamefont {P.~G.}\
  \bibnamefont {Tinyakov}},\ }\bibfield  {title} {\bibinfo {title} {{Instanton
  induced cross-sections below the sphaleron}},\ }\href
  {https://doi.org/10.1016/0550-3213(91)90267-2} {\bibfield  {journal}
  {\bibinfo  {journal} {Nucl. Phys.}\ }\textbf {\bibinfo {volume} {B350}},\
  \bibinfo {pages} {441} (\bibinfo {year} {1991})}\BibitemShut {NoStop}%
\bibitem [{\citenamefont {Coleman}\ and\ \citenamefont
  {De~Luccia}(1980)}]{Coleman:1980aw}%
  \BibitemOpen
  \bibfield  {author} {\bibinfo {author} {\bibfnamefont {S.~R.}\ \bibnamefont
  {Coleman}}\ and\ \bibinfo {author} {\bibfnamefont {F.}~\bibnamefont
  {De~Luccia}},\ }\bibfield  {title} {\bibinfo {title} {{Gravitational Effects
  on and of Vacuum Decay}},\ }\href {https://doi.org/10.1103/PhysRevD.21.3305}
  {\bibfield  {journal} {\bibinfo  {journal} {Phys. Rev. D}\ }\textbf {\bibinfo
  {volume} {21}},\ \bibinfo {pages} {3305} (\bibinfo {year}
  {1980})}\BibitemShut {NoStop}%
\bibitem [{\citenamefont {Hawking}\ and\ \citenamefont
  {Turok}(1998)}]{Hawking:1998bn}%
  \BibitemOpen
  \bibfield  {author} {\bibinfo {author} {\bibfnamefont {S.}~\bibnamefont
  {Hawking}}\ and\ \bibinfo {author} {\bibfnamefont {N.}~\bibnamefont
  {Turok}},\ }\bibfield  {title} {\bibinfo {title} {{Open inflation without
  false vacua}},\ }\href {https://doi.org/10.1016/S0370-2693(98)00234-2}
  {\bibfield  {journal} {\bibinfo  {journal} {Phys. Lett. B}\ }\textbf
  {\bibinfo {volume} {425}},\ \bibinfo {pages} {25} (\bibinfo {year} {1998})},\
  \Eprint {https://arxiv.org/abs/hep-th/9802030} {arXiv:hep-th/9802030}
  \BibitemShut {NoStop}%
\bibitem [{\citenamefont {Garriga}(2000)}]{Garriga:1998tm}%
  \BibitemOpen
  \bibfield  {author} {\bibinfo {author} {\bibfnamefont {J.}~\bibnamefont
  {Garriga}},\ }\bibfield  {title} {\bibinfo {title} {{Open inflation and the
  singular boundary}},\ }\href {https://doi.org/10.1103/PhysRevD.61.047301}
  {\bibfield  {journal} {\bibinfo  {journal} {Phys. Rev. D}\ }\textbf {\bibinfo
  {volume} {61}},\ \bibinfo {pages} {047301} (\bibinfo {year} {2000})},\
  \Eprint {https://arxiv.org/abs/hep-th/9803210} {arXiv:hep-th/9803210}
  \BibitemShut {NoStop}%
\bibitem [{\citenamefont {Vilenkin}(1998)}]{Vilenkin:1998pp}%
  \BibitemOpen
  \bibfield  {author} {\bibinfo {author} {\bibfnamefont {A.}~\bibnamefont
  {Vilenkin}},\ }\bibfield  {title} {\bibinfo {title} {{Singular instantons and
  creation of open universes}},\ }\href
  {https://doi.org/10.1103/PhysRevD.57.R7069} {\bibfield  {journal} {\bibinfo
  {journal} {Phys. Rev. D}\ }\textbf {\bibinfo {volume} {57}},\ \bibinfo
  {pages} {7069} (\bibinfo {year} {1998})},\ \Eprint
  {https://arxiv.org/abs/hep-th/9803084} {arXiv:hep-th/9803084} \BibitemShut
  {NoStop}%
\bibitem [{\citenamefont {Giddings}\ and\ \citenamefont
  {Strominger}(1988)}]{Giddings:1987cg}%
  \BibitemOpen
  \bibfield  {author} {\bibinfo {author} {\bibfnamefont {S.~B.}\ \bibnamefont
  {Giddings}}\ and\ \bibinfo {author} {\bibfnamefont {A.}~\bibnamefont
  {Strominger}},\ }\bibfield  {title} {\bibinfo {title} {{Axion Induced
  Topology Change in Quantum Gravity and String Theory}},\ }\href
  {https://doi.org/10.1016/0550-3213(88)90446-4} {\bibfield  {journal}
  {\bibinfo  {journal} {Nucl. Phys. B}\ }\textbf {\bibinfo {volume} {306}},\
  \bibinfo {pages} {890} (\bibinfo {year} {1988})}\BibitemShut {NoStop}%
\bibitem [{\citenamefont {Coleman}\ \emph {et~al.}(1978)\citenamefont
  {Coleman}, \citenamefont {Glaser},\ and\ \citenamefont
  {Martin}}]{Coleman:1977th}%
  \BibitemOpen
  \bibfield  {author} {\bibinfo {author} {\bibfnamefont {S.~R.}\ \bibnamefont
  {Coleman}}, \bibinfo {author} {\bibfnamefont {V.}~\bibnamefont {Glaser}},\
  and\ \bibinfo {author} {\bibfnamefont {A.}~\bibnamefont {Martin}},\
  }\bibfield  {title} {\bibinfo {title} {{Action Minima Among Solutions to a
  Class of Euclidean Scalar Field Equations}},\ }\href
  {https://doi.org/10.1007/BF01609421} {\bibfield  {journal} {\bibinfo
  {journal} {Commun. Math. Phys.}\ }\textbf {\bibinfo {volume} {58}},\ \bibinfo
  {pages} {211} (\bibinfo {year} {1978})}\BibitemShut {NoStop}%
\bibitem [{\citenamefont {Blum}\ \emph {et~al.}(2017)\citenamefont {Blum},
  \citenamefont {Honda}, \citenamefont {Sato}, \citenamefont {Takimoto},\ and\
  \citenamefont {Tobioka}}]{Blum:2016ipp}%
  \BibitemOpen
  \bibfield  {author} {\bibinfo {author} {\bibfnamefont {K.}~\bibnamefont
  {Blum}}, \bibinfo {author} {\bibfnamefont {M.}~\bibnamefont {Honda}},
  \bibinfo {author} {\bibfnamefont {R.}~\bibnamefont {Sato}}, \bibinfo {author}
  {\bibfnamefont {M.}~\bibnamefont {Takimoto}},\ and\ \bibinfo {author}
  {\bibfnamefont {K.}~\bibnamefont {Tobioka}},\ }\bibfield  {title} {\bibinfo
  {title} {{O($N$) Invariance of the Multi-Field Bounce}},\ }\href
  {https://doi.org/10.1007/JHEP05(2017)109, 10.1007/JHEP06(2017)060} {\bibfield
   {journal} {\bibinfo  {journal} {JHEP}\ }\textbf {\bibinfo {volume} {05}},\
  \bibinfo {pages} {109}},\ \bibinfo {note} {[Erratum: JHEP06,060(2017)]},\
  \Eprint {https://arxiv.org/abs/1611.04570} {arXiv:1611.04570 [hep-th]}
  \BibitemShut {NoStop}%
\bibitem [{\citenamefont {Adams}\ \emph {et~al.}(2006)\citenamefont {Adams},
  \citenamefont {Arkani-Hamed}, \citenamefont {Dubovsky}, \citenamefont
  {Nicolis},\ and\ \citenamefont {Rattazzi}}]{Adams:2006sv}%
  \BibitemOpen
  \bibfield  {author} {\bibinfo {author} {\bibfnamefont {A.}~\bibnamefont
  {Adams}}, \bibinfo {author} {\bibfnamefont {N.}~\bibnamefont {Arkani-Hamed}},
  \bibinfo {author} {\bibfnamefont {S.}~\bibnamefont {Dubovsky}}, \bibinfo
  {author} {\bibfnamefont {A.}~\bibnamefont {Nicolis}},\ and\ \bibinfo {author}
  {\bibfnamefont {R.}~\bibnamefont {Rattazzi}},\ }\bibfield  {title} {\bibinfo
  {title} {{Causality, analyticity and an IR obstruction to UV completion}},\
  }\href {https://doi.org/10.1088/1126-6708/2006/10/014} {\bibfield  {journal}
  {\bibinfo  {journal} {JHEP}\ }\textbf {\bibinfo {volume} {10}},\ \bibinfo
  {pages} {014}},\ \Eprint {https://arxiv.org/abs/hep-th/0602178}
  {arXiv:hep-th/0602178 [hep-th]} \BibitemShut {NoStop}%
\bibitem [{\citenamefont {Herrero-Valea}\ \emph {et~al.}()\citenamefont
  {Herrero-Valea}, \citenamefont {Timiryasov},\ and\ \citenamefont
  {Tokareva}}]{Herrero-Valea:2019hde}%
  \BibitemOpen
  \bibfield  {author} {\bibinfo {author} {\bibfnamefont {M.}~\bibnamefont
  {Herrero-Valea}}, \bibinfo {author} {\bibfnamefont {I.}~\bibnamefont
  {Timiryasov}},\ and\ \bibinfo {author} {\bibfnamefont {A.}~\bibnamefont
  {Tokareva}},\ }\bibfield  {title} {\bibinfo {title} {{To Positivity and
  Beyond, where Higgs-Dilaton Inflation has never gone before}}\ }\href
  {https://doi.org/10.1088/1475-7516/2019/11/042}
  {10.1088/1475-7516/2019/11/042},\ \Eprint {https://arxiv.org/abs/1905.08816}
  {arXiv:1905.08816 [hep-ph]} \BibitemShut {NoStop}%
\bibitem [{\citenamefont {Rubio}\ and\ \citenamefont
  {Tomberg}(2019)}]{Rubio:2019ypq}%
  \BibitemOpen
  \bibfield  {author} {\bibinfo {author} {\bibfnamefont {J.}~\bibnamefont
  {Rubio}}\ and\ \bibinfo {author} {\bibfnamefont {E.~S.}\ \bibnamefont
  {Tomberg}},\ }\bibfield  {title} {\bibinfo {title} {{Preheating in Palatini
  Higgs inflation}},\ }\href {https://doi.org/10.1088/1475-7516/2019/04/021}
  {\bibfield  {journal} {\bibinfo  {journal} {JCAP}\ }\textbf {\bibinfo
  {volume} {1904}}\bibfield  {number} {\bibinfo  {number} { (04)},\ \bibinfo
  {pages} {021}},\ }\Eprint {https://arxiv.org/abs/1902.10148}
  {arXiv:1902.10148 [hep-ph]} \BibitemShut {NoStop}%
\bibitem [{\citenamefont {Akrami}\ \emph {et~al.}()\citenamefont {Akrami} \emph
  {et~al.}}]{Akrami:2018odb}%
  \BibitemOpen
  \bibfield  {author} {\bibinfo {author} {\bibfnamefont {Y.}~\bibnamefont
  {Akrami}} \emph {et~al.} (\bibinfo {collaboration} {Planck}),\ }\bibfield
  {title} {\bibinfo {title} {{Planck 2018 results. X. Constraints on
  inflation}},\ }\href@noop {} {\ }\Eprint {https://arxiv.org/abs/1807.06211}
  {arXiv:1807.06211 [astro-ph.CO]} \BibitemShut {NoStop}%
\bibitem [{\citenamefont {Germani}\ and\ \citenamefont
  {Kehagias}(2010)}]{Germani:2010gm}%
  \BibitemOpen
  \bibfield  {author} {\bibinfo {author} {\bibfnamefont {C.}~\bibnamefont
  {Germani}}\ and\ \bibinfo {author} {\bibfnamefont {A.}~\bibnamefont
  {Kehagias}},\ }\bibfield  {title} {\bibinfo {title} {{New Model of Inflation
  with Non-minimal Derivative Coupling of Standard Model Higgs Boson to
  Gravity}},\ }\href {https://doi.org/10.1103/PhysRevLett.105.011302}
  {\bibfield  {journal} {\bibinfo  {journal} {Phys. Rev. Lett.}\ }\textbf
  {\bibinfo {volume} {105}},\ \bibinfo {pages} {011302} (\bibinfo {year}
  {2010})},\ \Eprint {https://arxiv.org/abs/1003.2635} {arXiv:1003.2635
  [hep-ph]} \BibitemShut {NoStop}%
\bibitem [{\citenamefont {Lerner}\ and\ \citenamefont
  {McDonald}(2010)}]{Lerner:2010mq}%
  \BibitemOpen
  \bibfield  {author} {\bibinfo {author} {\bibfnamefont {R.~N.}\ \bibnamefont
  {Lerner}}\ and\ \bibinfo {author} {\bibfnamefont {J.}~\bibnamefont
  {McDonald}},\ }\bibfield  {title} {\bibinfo {title} {{A Unitarity-Conserving
  Higgs Inflation Model}},\ }\href {https://doi.org/10.1103/PhysRevD.82.103525}
  {\bibfield  {journal} {\bibinfo  {journal} {Phys. Rev.}\ }\textbf {\bibinfo
  {volume} {D82}},\ \bibinfo {pages} {103525} (\bibinfo {year} {2010})},\
  \Eprint {https://arxiv.org/abs/1005.2978} {arXiv:1005.2978 [hep-ph]}
  \BibitemShut {NoStop}%
\bibitem [{\citenamefont {Giudice}\ and\ \citenamefont
  {Lee}(2011)}]{Giudice:2010ka}%
  \BibitemOpen
  \bibfield  {author} {\bibinfo {author} {\bibfnamefont {G.~F.}\ \bibnamefont
  {Giudice}}\ and\ \bibinfo {author} {\bibfnamefont {H.~M.}\ \bibnamefont
  {Lee}},\ }\bibfield  {title} {\bibinfo {title} {{Unitarizing Higgs
  Inflation}},\ }\href {https://doi.org/10.1016/j.physletb.2010.10.035}
  {\bibfield  {journal} {\bibinfo  {journal} {Phys. Lett.}\ }\textbf {\bibinfo
  {volume} {B694}},\ \bibinfo {pages} {294} (\bibinfo {year} {2011})},\ \Eprint
  {https://arxiv.org/abs/1010.1417} {arXiv:1010.1417 [hep-ph]} \BibitemShut
  {NoStop}%
\bibitem [{\citenamefont {Hoang}(2014)}]{Hoang:2014oea}%
  \BibitemOpen
  \bibfield  {author} {\bibinfo {author} {\bibfnamefont {A.~H.}\ \bibnamefont
  {Hoang}},\ }\bibfield  {title} {\bibinfo {title} {{The Top Mass:
  Interpretation and Theoretical Uncertainties}},\ }in\ \href@noop {} {\emph
  {\bibinfo {booktitle} {{Proceedings, 7th International Workshop on Top Quark
  Physics (TOP2014): Cannes, France, September 28-October 3, 2014}}}}\
  (\bibinfo {year} {2014})\ \Eprint {https://arxiv.org/abs/1412.3649}
  {arXiv:1412.3649 [hep-ph]} \BibitemShut {NoStop}%
\bibitem [{\citenamefont {Khachatryan}\ \emph {et~al.}(2016)\citenamefont
  {Khachatryan} \emph {et~al.}}]{Khachatryan:2015hba}%
  \BibitemOpen
  \bibfield  {author} {\bibinfo {author} {\bibfnamefont {V.}~\bibnamefont
  {Khachatryan}} \emph {et~al.} (\bibinfo {collaboration} {CMS}),\ }\bibfield
  {title} {\bibinfo {title} {{Measurement of the top quark mass using
  proton-proton data at ${\sqrt{(s)}}$ = 7 and 8 TeV}},\ }\href
  {https://doi.org/10.1103/PhysRevD.93.072004} {\bibfield  {journal} {\bibinfo
  {journal} {Phys. Rev.}\ }\textbf {\bibinfo {volume} {D93}},\ \bibinfo {pages}
  {072004} (\bibinfo {year} {2016})},\ \Eprint
  {https://arxiv.org/abs/1509.04044} {arXiv:1509.04044 [hep-ex]} \BibitemShut
  {NoStop}%
\bibitem [{\citenamefont {Ferrario~Ravasio}\ \emph {et~al.}(2018)\citenamefont
  {Ferrario~Ravasio}, \citenamefont {Ježo}, \citenamefont {Nason},\ and\
  \citenamefont {Oleari}}]{Ravasio:2018lzi}%
  \BibitemOpen
  \bibfield  {author} {\bibinfo {author} {\bibfnamefont {S.}~\bibnamefont
  {Ferrario~Ravasio}}, \bibinfo {author} {\bibfnamefont {T.}~\bibnamefont
  {Ježo}}, \bibinfo {author} {\bibfnamefont {P.}~\bibnamefont {Nason}},\ and\
  \bibinfo {author} {\bibfnamefont {C.}~\bibnamefont {Oleari}},\ }\bibfield
  {title} {\bibinfo {title} {{A theoretical study of top-mass measurements at
  the LHC using NLO+PS generators of increasing accuracy}},\ }\href
  {https://doi.org/10.1140/epjc/s10052-018-5909-7} {\bibfield  {journal}
  {\bibinfo  {journal} {Eur. Phys. J.}\ }\textbf {\bibinfo {volume} {C78}},\
  \bibinfo {pages} {458} (\bibinfo {year} {2018})},\ \Eprint
  {https://arxiv.org/abs/1801.03944} {arXiv:1801.03944 [hep-ph]} \BibitemShut
  {NoStop}%
\bibitem [{\citenamefont {Aaboud}\ \emph {et~al.}(2019)\citenamefont {Aaboud}
  \emph {et~al.}}]{Aaboud:2018zbu}%
  \BibitemOpen
  \bibfield  {author} {\bibinfo {author} {\bibfnamefont {M.}~\bibnamefont
  {Aaboud}} \emph {et~al.} (\bibinfo {collaboration} {ATLAS}),\ }\bibfield
  {title} {\bibinfo {title} {{Measurement of the top quark mass in the
  $t\bar{t}\rightarrow $ lepton+jets channel from $\sqrt{s}=8$ TeV ATLAS data
  and combination with previous results}},\ }\href
  {https://doi.org/10.1140/epjc/s10052-019-6757-9} {\bibfield  {journal}
  {\bibinfo  {journal} {Eur. Phys. J.}\ }\textbf {\bibinfo {volume} {C79}},\
  \bibinfo {pages} {290} (\bibinfo {year} {2019})},\ \Eprint
  {https://arxiv.org/abs/1810.01772} {arXiv:1810.01772 [hep-ex]} \BibitemShut
  {NoStop}%
\bibitem [{\citenamefont {Rasanen}\ and\ \citenamefont
  {Wahlman}(2017)}]{Rasanen:2017ivk}%
  \BibitemOpen
  \bibfield  {author} {\bibinfo {author} {\bibfnamefont {S.}~\bibnamefont
  {Rasanen}}\ and\ \bibinfo {author} {\bibfnamefont {P.}~\bibnamefont
  {Wahlman}},\ }\bibfield  {title} {\bibinfo {title} {{Higgs inflation with
  loop corrections in the Palatini formulation}},\ }\href
  {https://doi.org/10.1088/1475-7516/2017/11/047} {\bibfield  {journal}
  {\bibinfo  {journal} {JCAP}\ }\textbf {\bibinfo {volume} {1711}}\bibfield
  {number} {\bibinfo  {number} { (11)},\ \bibinfo {pages} {047}},\ }\Eprint
  {https://arxiv.org/abs/1709.07853} {arXiv:1709.07853 [astro-ph.CO]}
  \BibitemShut {NoStop}%
\bibitem [{\citenamefont {Bezrukov}\ and\ \citenamefont
  {Shaposhnikov}(2009)}]{Bezrukov:2009db}%
  \BibitemOpen
  \bibfield  {author} {\bibinfo {author} {\bibfnamefont {F.}~\bibnamefont
  {Bezrukov}}\ and\ \bibinfo {author} {\bibfnamefont {M.}~\bibnamefont
  {Shaposhnikov}},\ }\bibfield  {title} {\bibinfo {title} {{Standard Model
  Higgs boson mass from inflation: Two loop analysis}},\ }\href
  {https://doi.org/10.1088/1126-6708/2009/07/089} {\bibfield  {journal}
  {\bibinfo  {journal} {JHEP}\ }\textbf {\bibinfo {volume} {07}},\ \bibinfo
  {pages} {089}},\ \Eprint {https://arxiv.org/abs/0904.1537} {arXiv:0904.1537
  [hep-ph]} \BibitemShut {NoStop}%
\bibitem [{\citenamefont {Bezrukov}\ \emph {et~al.}(2018)\citenamefont
  {Bezrukov}, \citenamefont {Pauly},\ and\ \citenamefont
  {Rubio}}]{Bezrukov:2017dyv}%
  \BibitemOpen
  \bibfield  {author} {\bibinfo {author} {\bibfnamefont {F.}~\bibnamefont
  {Bezrukov}}, \bibinfo {author} {\bibfnamefont {M.}~\bibnamefont {Pauly}},\
  and\ \bibinfo {author} {\bibfnamefont {J.}~\bibnamefont {Rubio}},\ }\bibfield
   {title} {\bibinfo {title} {{On the robustness of the primordial power
  spectrum in renormalized Higgs inflation}},\ }\href
  {https://doi.org/10.1088/1475-7516/2018/02/040} {\bibfield  {journal}
  {\bibinfo  {journal} {JCAP}\ }\textbf {\bibinfo {volume} {02}},\ \bibinfo
  {pages} {040}},\ \Eprint {https://arxiv.org/abs/1706.05007} {arXiv:1706.05007
  [hep-ph]} \BibitemShut {NoStop}%
\bibitem [{\citenamefont {Bezrukov}\ \emph {et~al.}(2009)\citenamefont
  {Bezrukov}, \citenamefont {Magnin},\ and\ \citenamefont
  {Shaposhnikov}}]{Bezrukov_2009}%
  \BibitemOpen
  \bibfield  {author} {\bibinfo {author} {\bibfnamefont {F.~L.}\ \bibnamefont
  {Bezrukov}}, \bibinfo {author} {\bibfnamefont {A.}~\bibnamefont {Magnin}},\
  and\ \bibinfo {author} {\bibfnamefont {M.}~\bibnamefont {Shaposhnikov}},\
  }\bibfield  {title} {\bibinfo {title} {Standard model higgs boson mass from
  inflation},\ }\href {https://doi.org/10.1016/j.physletb.2009.03.035}
  {\bibfield  {journal} {\bibinfo  {journal} {Physics Letters B}\ }\textbf
  {\bibinfo {volume} {675}},\ \bibinfo {pages} {88?92} (\bibinfo {year}
  {2009})}\BibitemShut {NoStop}%
\bibitem [{\citenamefont {Falls}\ and\ \citenamefont
  {Herrero-Valea}(2019)}]{Falls:2018olk}%
  \BibitemOpen
  \bibfield  {author} {\bibinfo {author} {\bibfnamefont {K.}~\bibnamefont
  {Falls}}\ and\ \bibinfo {author} {\bibfnamefont {M.}~\bibnamefont
  {Herrero-Valea}},\ }\bibfield  {title} {\bibinfo {title} {{Frame
  (In)equivalence in Quantum Field Theory and Cosmology}},\ }\href
  {https://doi.org/10.1140/epjc/s10052-019-7070-3} {\bibfield  {journal}
  {\bibinfo  {journal} {Eur. Phys. J. C}\ }\textbf {\bibinfo {volume} {79}},\
  \bibinfo {pages} {595} (\bibinfo {year} {2019})},\ \Eprint
  {https://arxiv.org/abs/1812.08187} {arXiv:1812.08187 [hep-th]} \BibitemShut
  {NoStop}%
\bibitem [{\citenamefont {Fradkin}\ and\ \citenamefont
  {Vilkovisky}(1973)}]{Fradkin:1974df}%
  \BibitemOpen
  \bibfield  {author} {\bibinfo {author} {\bibfnamefont {E.}~\bibnamefont
  {Fradkin}}\ and\ \bibinfo {author} {\bibfnamefont {G.}~\bibnamefont
  {Vilkovisky}},\ }\bibfield  {title} {\bibinfo {title} {{S matrix for
  gravitational field. ii. local measure, general relations, elements of
  renormalization theory}},\ }\href {https://doi.org/10.1103/PhysRevD.8.4241}
  {\bibfield  {journal} {\bibinfo  {journal} {Phys. Rev. D}\ }\textbf {\bibinfo
  {volume} {8}},\ \bibinfo {pages} {4241} (\bibinfo {year} {1973})}\BibitemShut
  {NoStop}%
\bibitem [{\citenamefont {Henneaux}\ and\ \citenamefont
  {Teitelboim}(1992)}]{Henneaux:1992ig}%
  \BibitemOpen
  \bibfield  {author} {\bibinfo {author} {\bibfnamefont {M.}~\bibnamefont
  {Henneaux}}\ and\ \bibinfo {author} {\bibfnamefont {C.}~\bibnamefont
  {Teitelboim}},\ }\href@noop {} {\emph {\bibinfo {title} {{Quantization of
  gauge systems}}}}\ (\bibinfo {year} {1992})\BibitemShut {NoStop}%
\bibitem [{\citenamefont {Lee}\ and\ \citenamefont {Yang}(1962)}]{Lee:1962vm}%
  \BibitemOpen
  \bibfield  {author} {\bibinfo {author} {\bibfnamefont {T.}~\bibnamefont
  {Lee}}\ and\ \bibinfo {author} {\bibfnamefont {C.-N.}\ \bibnamefont {Yang}},\
  }\bibfield  {title} {\bibinfo {title} {{Theory of Charged Vector Mesons
  Interacting with the Electromagnetic Field}},\ }\href
  {https://doi.org/10.1103/PhysRev.128.885} {\bibfield  {journal} {\bibinfo
  {journal} {Phys. Rev.}\ }\textbf {\bibinfo {volume} {128}},\ \bibinfo {pages}
  {885} (\bibinfo {year} {1962})}\BibitemShut {NoStop}%
\bibitem [{\citenamefont {Salam}\ and\ \citenamefont
  {Strathdee}(1970)}]{Salam:1971sp}%
  \BibitemOpen
  \bibfield  {author} {\bibinfo {author} {\bibfnamefont {A.}~\bibnamefont
  {Salam}}\ and\ \bibinfo {author} {\bibfnamefont {J.}~\bibnamefont
  {Strathdee}},\ }\bibfield  {title} {\bibinfo {title} {{Equivalent
  formulations of massive vector field theories}},\ }\href
  {https://doi.org/10.1103/PhysRevD.2.2869} {\bibfield  {journal} {\bibinfo
  {journal} {Phys. Rev. D}\ }\textbf {\bibinfo {volume} {2}},\ \bibinfo {pages}
  {2869} (\bibinfo {year} {1970})}\BibitemShut {NoStop}%
\bibitem [{\citenamefont {Gerstein}\ \emph {et~al.}(1971)\citenamefont
  {Gerstein}, \citenamefont {Jackiw}, \citenamefont {Weinberg},\ and\
  \citenamefont {Lee}}]{Gerstein:1971fm}%
  \BibitemOpen
  \bibfield  {author} {\bibinfo {author} {\bibfnamefont {I.}~\bibnamefont
  {Gerstein}}, \bibinfo {author} {\bibfnamefont {R.}~\bibnamefont {Jackiw}},
  \bibinfo {author} {\bibfnamefont {S.}~\bibnamefont {Weinberg}},\ and\
  \bibinfo {author} {\bibfnamefont {B.}~\bibnamefont {Lee}},\ }\bibfield
  {title} {\bibinfo {title} {{Chiral loops}},\ }\href
  {https://doi.org/10.1103/PhysRevD.3.2486} {\bibfield  {journal} {\bibinfo
  {journal} {Phys. Rev. D}\ }\textbf {\bibinfo {volume} {3}},\ \bibinfo {pages}
  {2486} (\bibinfo {year} {1971})}\BibitemShut {NoStop}%
\bibitem [{\citenamefont {Aros}\ \emph {et~al.}(2003)\citenamefont {Aros},
  \citenamefont {Contreras},\ and\ \citenamefont {Zanelli}}]{Aros:2003bi}%
  \BibitemOpen
  \bibfield  {author} {\bibinfo {author} {\bibfnamefont {R.}~\bibnamefont
  {Aros}}, \bibinfo {author} {\bibfnamefont {M.}~\bibnamefont {Contreras}},\
  and\ \bibinfo {author} {\bibfnamefont {J.}~\bibnamefont {Zanelli}},\
  }\bibfield  {title} {\bibinfo {title} {{Path integral measure for first order
  and metric gravities}},\ }\href {https://doi.org/10.1088/0264-9381/20/13/336}
  {\bibfield  {journal} {\bibinfo  {journal} {Class. Quant. Grav.}\ }\textbf
  {\bibinfo {volume} {20}},\ \bibinfo {pages} {2937} (\bibinfo {year}
  {2003})},\ \Eprint {https://arxiv.org/abs/gr-qc/0303113}
  {arXiv:gr-qc/0303113} \BibitemShut {NoStop}%
\end{thebibliography}%

\end{document}